\documentclass[preprint,showpacs,preprintnumbers,amsmath,amssymb,endfloa
ts*]{revtex4}
\usepackage{graphicx}

\begin{document}
\def \ee {\varepsilon}
\def \ef {{\varepsilon^{(1)}}}
\def \es {{\varepsilon^{(2)}}}
\def \ezf {{\varepsilon_0^{(1)}}}
\def \ezs {{\varepsilon_0^{(2)}}}
\def \rf {{r_0^{(1)}}}
\def \rs {{r_0^{(2)}}}

\thispagestyle{empty}
\title{
Thermal quantum field theory and the Casimir interaction
between dielectrics
}
\author{
B.~Geyer,
G.~L.~Klimchitskaya,\footnote{
On leave from
North-West Technical University,\\
St.Petersburg, Russia.}
and V.~M.~Mostepanenko\footnote{
On leave from
Noncommercial Partnership  ``Scientific\\ Instruments'' Moscow, Russia.}
}

\affiliation{
Center of Theoretical Studies and Institute for Theoretical Physics,\\
Leipzig University, Augustusplatz 10/11, D-04109 Leipzig, Germany
}

\begin{abstract}
The Casimir and van der Waals interaction between two dissimilar thick
dielectric plates is reconsidered on the basis of thermal quantum field
theory in Matsubara formulation. We briefly review two main
derivations of the Lifshitz formula in the framework of thermal
quantum field theory without use of the fluctuation-dissipation
theorem. A set of special conditions is formulated under which these
derivations remain valid in the presence of dissipation.
The low-temperature behavior of the Casimir and van der Waals
interactions between dissimilar dielectrics is found analytically
from the Lifshitz theory for both an idealized model of dilute
dielectrics and for real dielectrics with finite static dielectric
permittivities. The free energy, pressure and entropy of the
Casimir and van der Waals interactions at low temperatures
demonstrate the same universal dependence on the temperature as
was previously discovered for ideal metals. The entropy vanishes
when temperature goes to zero proving the validity of the Nernst
heat theorem. This solves the long-standing problem on the
consistency of the Lifshitz theory with thermodynamics in the
case of dielectric plates. The obtained asymptotic expressions
are compared with numerical computations for both dissimilar
and similar real dielectrics and found to be in excellent
agreement. The role of the zero-frequency term in Matsubara sum
is investigated  in the case of dielectric plates. It is shown that
the inclusion of conductivity in the model
of dielectric response leads to the violation of the Nernst heat
theorem. The applications of this result to the topical problems
of noncontact atomic friction and the Casimir interaction between
real metals are discussed.
\end{abstract}
\maketitle

\section{Introduction}

Both the Casimir and van der Waals interactions are the quantum
phenomena caused by fluctuating electromagnetic fields.
In the framework of quantum field theory, these interactions
can be described through the alteration by the material
boundaries of the zero-point electromagnetic energy that
pervades all of space. Thus, the Casimir and
van der Waals forces are found to be closely connected with the
fundamental properties of quantum vacuum (see the original
Ref.~\cite{1} and monographs \cite{2,M_T,M_N,5}).

In the last few years the Casimir interaction has been actively
investigated in connection with topical applications
in extra-dimensional physics (where it provides an effective
mechanism for spontaneous compactification of extra spatial
dimensions) and in the bag model of hadrons \cite{M_T,B_M_M}.
Several measurements of the Casimir and van der Waals forces
between metal macrobodies were performed with increased
precision using modern laboratory techniques
\cite{7,8,9,10,11,12,13,14,Decca1,Decca2,17,Si}.
The experimental results were used to obtain stronger constraints
on the Yukawa-type long-range interaction, predicted by many
extensions to the Standard Model, in the micrometer interaction range
\cite{Decca1,Decca2,17,18,19,20,21,22,23,24}.
Concurrent with the fundamental applications in elementary particle
physics, the use of the Casimir and van der Waals
forces in nanotechnology \cite{25,26,27}, in quantum reflection
and Bose-Einstein condensation \cite{28,29} and in noncontact
atomic friction \cite{30,atFric,32} was begun.

The numerous applications of the Casimir and van der Waals
forces and the extensive experimental work impose severe demands
on the accuracy of theoretical predictions. The basic theory of
these forces acting between real materials was developed by
E.~M.~Lifshitz and his collaborators \cite{Lifsh,D_L_P,L_P}.
It expresses the free energy and force acting between two
macrobodies in terms of their frequency-dependent dielectric
permittivities. The original derivation of the Lifshitz formulas
for the free energy and pressure between two thick dielectric
plates is based on the concept of a fluctuating electromagnetic
field and uses the fluctuation-dissipation theorem. If the
temperature of plates is not equal to zero, this concept includes
both zero-point oscillations and thermal fluctuations of
the field. In general, the Lifshitz derivation is applicable to
both transparent and absorbing media. At present time,
there are derivations of the Lifshitz formula based on the
thermal quantum field theory in the Matsubara formulation (see
Ref.~\cite{B_M_M} for a review). They do not employ the
fluctuation-dissipation theorem but lead to results identical
to those obtained with the use of this theorem. Thermal
quantum field theory is immediately applicable to only transparent
bodies, but, under certain conditions, dissipation can also be
included.

Quite independently from the Lifshitz formula, the Matsubara thermal
quantum field theory was applied \cite{Mehra,Maclay} for the
investigation of the Casimir force acting between two plane
parallel plates made of ideal metal (at zero temperature this
case was considered already in Ref.~\cite{1}). This is an exactly
solvable problem of field quantization with the Dirichlet boundary
conditions on the surface of the plates
for the tangential component of electric field  and with the
identification
condition in the Euclidean time variable. The obtained
analytic expressions for the Casimir pressure, free energy and
entropy were found to be in perfect agreement with
thermodynamics. In particular, the magnitudes of the free
energy and pressure were found to be monotonously
increasing functions with the increase of the temperature.
Entropy is nonnegative and goes to zero when temperature
vanishes; this is in accordance with the third law of thermodynamics
(the Nernst heat theorem) \cite{Robash}.
For the first time, the Lifshitz
formula was considered to be in contradiction with the quantum
field theoretical approach to the case of ideal metals since it
leads to a different result in the limit of infinitely large
dielectric permittivity. The situation was clarified by the
so called Schwinger's prescription \cite{5,39} whereby the
limit of infinite dielectric permittivity in the Lifshitz
formula must be taken prior to putting the Matsubara frequency
equal to zero. Once that prescription is followed, the
results obtained for ideal metals from the Lifshitz formula
match those obtained
from thermal quantum field theory with the Dirichlet
boundary conditions.

Currently the Lifshitz formula leads to problems when applied to
the case of two plates made of real metals at nonzero
temperature.  In Refs.~\cite{B_S,41,42,43} metallic plates
were characterized with the dielectric permittivity of the Drude
model. In the limit of ideal metals this approach comes into
conflict with the values of the Casimir free energy and
pressure obtained using thermal quantum field theory with
the Dirichlet boundary conditions. As was also proved in
Refs.~\cite{R1,R4}, the substitution of the Drude dielectric
function in the Lifshitz formula leads to a violation of the
third law of thermodynamics in the case of perfect crystal
lattice with
no defects and impurities (as discussed above, the quantum
field theory approach obeys this law in the case of ideal metals).

Another way to describe the realistic properties of a metal
is to use the dielectric permittivity of the free electron
plasma model \cite{39,46,R2} or to impose the surface-impedance
boundary conditions \cite{R4,GKM_03}.
In the case of ideal metals, both these approaches
are in agreement with the results
obtained from using thermal quantum field theory.
They also obey the third law of
thermodynamics. Recent experiments permit to test different
theoretical approaches to the thermal Casimir force. It is
significant that the first modern measurement of the Casimir
force \cite{7} was found \cite{Lam05,Lam_T} to be in
disagreement with Refs.~\cite{B_S,41,42,43} and consistent
with Ref.~\cite{GKM_03}. The most precise and accurate
experiments of Refs.~\cite{Decca1,Decca2,17} exclude the
theoretical approach of Refs.~\cite{B_S,41,42,43} at 99\%
confidence and are consistent with Refs.~\cite{R4,46,R2,GKM_03}.
(Note that there is some controversy in the literature on the
agreement of different approaches with thermodynamics and
experiment which is reflected in recent Refs.~\cite{51,52}.)

As is evident from the foregoing, there are many
questions in the Lifshitz theory of the Casimir and
van der Waals interactions that remain to be answered. On the
one hand, the consistency of the Lifshitz theory with
thermal quantum field theory is not completely
understood, and, on the other hand, the agreement of the
Lifshitz theory with thermodynamics is called into question.
An important point is that some thermodynamic aspects
of the Lifshitz theory are still unknown even in the case
of two dielectric plates for which consensus in the literature
is achieved. One of the major problems realized 50 years back,
but not resolved up to the present, is the elucidation
of the low-temperature behavior of the Casimir (van der Waals)
free energy, pressure and entropy between two dielectric
plates. Without a resolution of this fundamental issue it
would be impossible to settle the more complicated problems
related to real metals. Even the correct way of practical
computations for comparison of experimental results with
theory would be uncertain.

In the present paper we reconsider the thermal van der Waals
and Casimir interactions between two thick dissimilar dielectric
plates. We start with a brief analysis of two derivations of
the Lifshitz formula in the framework of thermal quantum
field theory in the Matsubara formulation. Special attention
is paid to the conditions under which the quantum field
theoretical derivations are applicable not only to transparent
media but also to media with dissipation. We next consider the
Casimir effect between two plates made of dissimilar dilute
dielectrics with constant dielectric permittivities
${\ee}^{(k)}$ ($k=1,\,2$). In this case it is possible to
develop the perturbation theory in the small parameters
${\ee}^{(k)}-1$  and to obtain the explicit analytic expressions
for the Casimir free energy and pressure which are exact
in the separation between the plates and their temperature.
The simple asymptotic behavior of the free energy, pressure and
entropy at both low and high temperatures (short and large
separations, respectively) is also found. The leading terms
of the Casimir entropy at low temperature are shown to be
of the second and third power in temperature, thus,
demonstrating the agreement of the Lifshitz formula with
thermodynamics in the case of dilute dielectrics.

As a next step, we derive analytically the low-temperature
behavior of the Casimir and van der Waals forces between dissimilar
dielectrics using the frequency-dependent dielectric permittivities.
For this purpose, perturbation theory in a small parameter
which is proportional to the product of the separation
distance between the plates and their temperature is developed.
It is proved that
the thermal corrections to the van der Waals and Casimir energy and
pressure have an universal form and their asymptotic behavior can
be calculated in terms of the static dielectric permittivities of
the plate materials. Once again, the leading contributions to the
entropy are shown to be of the second and third power in
temprature which proves the agreement of the Lifshitz formula with
thermodynamics in the case of dielectric plates with finite
static dielectric permittivities. In the limit
${\ee}^{(k)}-1\ll 1$ the above results for dilute dielectrics are
again obtained. Results for the special case of similar dielectrics
are also provided. This means that the long-standing problem
on the low-temperature properties of the Lifshitz theory in the
case of dielectric plates is resolved.

In what follows we perform the comparison between the obtained
analytic results and numerical computations for dielectric plates
made of silicon and vitreous silica. The cases of both dissimilar
and similar plates are considered. The dielectric permittivities
of silicon and vitreous silica along the imaginary frequency axis
are found by means of the dispersion relation using the tabulated
optical data for both materials. In all cases the
excellent agreement between the analytic and numerical
results is observed below some definite temperature
(separation) values.

Finally we discuss the role of the zero-frequency term in the
Matsubara sum for dielectrics. To calculate the van der Waals
friction, Refs.~\cite{atFric,SurfSci} use the Lifshitz-type
formula and describe the dielectric permittivity of the dielectric
substrate by means of the Drude model with appropriately low
conductivity. Formally, this leads to infinitely high
permittivity of a dielectric at zero frequency and to the
modification of the zero-frequency term of the Lifshitz
formula. We demonstrate that such a modification results in the
violation of the Nernst heat theorem. Hence, it follows that
the dc conductivity of dielectrics is
irrelevant to the van der Waals and Casimir forces and must
not be included in the model of dielectric response.
This conclusion leads to important consequences for both
the problem of atomic friction and for the Casimir and
van der Waals interactions between real metals.

The paper is organized as follows. In Sec.~II, two main
derivations of the Lifshitz formula for the free energy in
the framework of thermal quantum field theory are
briefly discussed. Special attention is paid to the
restrictions imposed on the type of relaxation. Sec.~III
is devoted to the case of dissimilar dilute dielectrics.
In Sec.~IV, we derive the low-temperature behavior of the
van der Waals and Casimir interactions between dissimilar
dielectrics with frequency-dependent dielectric
permittivities.
Here, the consistency of the Lifshitz theory
with thermodynamic requirements is proved. In Sec.~V the
obtained analytic results are compared with the results of
numerical computations performed for some real dielectrics.
Sec.~VI demonstrates that the inclusion of the conductivity
of dielectrics in the model of dielectric
response leads to the violation of the Nernst heat theorem.
In Sec.~VII we present our conclusions and discussion
touching on the topical problems of the noncontact atomic friction
and the Casimir interaction between real metals.
Appendices A, B and C contain the mathematical proofs of some
statements used in Secs.~IV and VI.

\section{Derivations of the Lifshitz formula in Matsubara
quantum field theory}

We consider two thick dielectric plates (semispaces) with
the frequency-dependent dielectric permittivities
${\ee}^{(1)}(\omega)$ and ${\ee}^{(2)}(\omega)$,
restricted by the two parallel planes $z=\pm a/2$
with separation $a$ between them, in
thermal equilibrium at temperature $T$.

The original Lifshitz derivation of his formula \cite{Lifsh}
was based on the assumption that the dielectric materials are
characterized by randomly fluctuating sources of
long wavelength electromagnetic fields. This concept
includes not only the thermal fluctuations but also the
zero-point oscillations of the field \cite{L_P}.
Lifshitz based his derivation on the fluctuation-dissipation
theorem using the properties of electromagnetic
fluctuations.

There are two derivations of the Lifshitz formula at
nonzero temperature in the framework of thermal quantum field
theory without use of the fluctuation-dissipation
theorem. The first of them is due to M.~Bordag \cite{B_M_M}
and is based on the scattering approach \cite{scatter}.
An electromagnetic wave $e^{ikz}$ coming from the negative
part of $z$-axis in the dielectric with the permittivity
${\ee}^{(1)}(\omega)$ will be scattered on the empty gap
between the two semispaces and there occurs a transmitted
and a reflected wave,
\begin{eqnarray}
&&
\begin{array}{lcr}
\varphi_{1}(z)  &  \raisebox{-0.9ex}{${\sim\atop z\to -\infty}$} &
e^{ikz}+s_{12}e^{-ikz},
\end{array}
\nonumber \\
&&
\begin{array}{lcr}
\varphi_{1}(z) &
\raisebox{-0.9ex}{${\sim\atop z\to
    \infty}$} & s_{11}e^{ikz}.
\end{array}
\label{eqI1}
\end{eqnarray}
\noindent
A similar situation holds for the linear independent wave
$e^{-ikz}$ coming from the positive part of $z$-axis
\begin{eqnarray}
&&
\begin{array}{lcr}
\varphi_{2}(z)  &  \raisebox{-0.9ex}{${\sim\atop z\to -\infty}$} &
s_{22}e^{-ikz},
\end{array}
\nonumber \\
&&
\begin{array}{lcr}
\varphi_{2}(z)& \raisebox{-0.9ex}{${\sim\atop z\to
    \infty}$}  & e^{-ikz} +s_{21}  e^{ikz} \,.
\end{array}
\label{eqI2}
\end{eqnarray}
\noindent
The matrix $\{s_{ij}\}$ composed of the coefficients in
Eqs.~(\ref{eqI1}) and (\ref{eqI2}) is unitary.

For the evaluation of the free energy one has to use the
Euclidean version of the field theory obtained by Wick
rotation $x_0\to ix_4$ with the electromagnetic field periodic
in the Euclidean time variable with a period
$\beta=\hbar/(k_BT)$, where $k_B$ is the Boltzmann constant.
Starting from the usual representation of the free energy
${\cal F}$ in thermal quantum field theory at the one loop
level, we arrive at the result (see Ref.~\cite{B_M_M} for
details)
\begin{eqnarray}
&&
{\cal F}(a,T)=-\frac{\hbar}{\beta}
\sum\limits_{l=0}^{\infty}
\left(1-\frac{1}{2}\delta_{l0}\right)
\int_{0}^{\infty}\frac{k_{\bot}dk_{\bot}}{2\pi}
\nonumber \\
&&\phantom{aa}
\times
\left[\ln s_{11}^{\|}(i\xi_l,k_{\bot})+
\ln s_{11}^{\bot}(i\xi_l,k_{\bot})\right].
\label{eqI3}
\end{eqnarray}
\noindent
Here $s_{11}^{\|,\bot}$ are the elements of the scattering
matrix for the two independent polarizations of electromagnetic
field, $k_{\bot}=|{\bf k}_{\bot}|$ is the magnitude of the
wave vector in the plane of plates, and
$\xi_l=2\pi k_BTl/\hbar$ are the Matsubara frequencies.

To obtain the explicit expressions for the elements of the
scattering matrix, we assume that the electric field satisfy
the Maxwell equations and the corresponding boundary
conditions on the
boundary surfaces $z=\pm a/2$ (i.e., that the normal
components of ${\bf D}^{(k)}={\ee}^{(k)}{\bf E}^{(k)}$
and tangential components of ${\bf E}^{(k)}$ are continuous).
By solving the scattering problem with the demand that
for infinitely remote plates both the free energy and pressure
are equal to zero, one obtains \cite{B_M_M}
\begin{eqnarray}
&&
s_{11}^{\|}(i\xi_l,k_{\bot})=\left(1-
\frac{{\ee}_{l}^{(1)}q_l-k_l^{(1)}}{{\ee}_{l}^{(1)}q_l+k_l^{(1)}}
\,
\frac{{\ee}_{l}^{(2)}q_l-k_l^{(2)}}{{\ee}_{l}^{(2)}q_l+k_l^{(2)}}
\,e^{-2aq_l}\right)^{-1},
\nonumber \\
&&
s_{11}^{\bot}(i\xi_l,k_{\bot})=\left(1-
\frac{k_l^{(1)}-q_l}{k_l^{(1)}+q_l}\,
\frac{k_l^{(2)}-q_l}{k_l^{(2)}+q_l}\,
e^{-2aq_l}\right)^{-1},
\label{eqI4}
\end{eqnarray}
where
\begin{equation}
q_l=\sqrt{\frac{\xi_l^2}{c^2}+k_{\bot}^2},
\quad
k_l^{(k)}=\sqrt{{\ee}^{(k)}(i\xi_l)\frac{\xi_l^2}{c^2}
+k_{\bot}^2}.
\label{eqI5}
\end{equation}
\noindent
An important point is that the Maxwell equations supplemented
by the corresponding boundary
conditions lead to a definite solution of the
scattering problem, given by Eqs.~(\ref{eqI4}) and (\ref{eqI5}),
only for $l\geq 1$. For $l=0$, however, $q_0=k_0^{(k)}=k_{\bot}$
and, as a result, the respective system of linear algebraic
equations has infinitely many solutions, i.e., it is
satisfied for any $s_{11}^{\bot}(0,k_{\bot})$ \cite{GKM_01}.
The definite value of  $s_{11}^{\bot}(0,k_{\bot})$ is obtained
by the use of the unitarity condition resulting in
$|s_{11}^{\bot}(0,k_{\bot})|=1$ and dispersion relation
leading to $s_{11}^{\bot}(0,k_{\bot})=1$. We will return
to this point below when discussing the role of dissipation.

Substituting Eq.~(\ref{eqI4}) in Eq.~(\ref{eqI3}), we arrive at
the Lifshitz formula for the free energy of the van der Waals
and Casimir interaction
\begin{eqnarray}
&&
{\cal F}(a,T)=\frac{k_BT}{2\pi}
\sum\limits_{l=0}^{\infty}
\left(1-\frac{1}{2}\delta_{l0}\right)
\int_{0}^{\infty}k_{\bot}dk_{\bot}
\nonumber \\
&&\phantom{aa}
\times
\left\{\ln\left[1-r_{\|}^{(1)}(\xi_l,k_{\bot})
r_{\|}^{(2)}(\xi_l,k_{\bot})e^{-2aq_l}\right]\right.
\label{eqI6} \\
&&\phantom{aaa}
\left.+
\ln\left[1-r_{\bot}^{(1)}(\xi_l,k_{\bot})
r_{\bot}^{(2)}(\xi_l,k_{\bot})e^{-2aq_l}\right]
\right\},
\nonumber
\end{eqnarray}
\noindent
where the reflection coefficients are defined by
\begin{eqnarray}
&&
r_{\|}^{(k)}(\xi_l,k_{\bot})=
\frac{{\ee}_l^{(k)}q_l-k_l^{(k)}}{{\ee}_l^{(k)}q_l+k_l^{(k)}},
\nonumber \\
&&
r_{\bot}^{(k)}(\xi_l,k_{\bot})=
\frac{k_l^{(k)}-q_l}{k_l^{(k)}+q_l}.
\label{eqI7}
\end{eqnarray}

The second field theoretical approach to the derivation of
the Lifshitz formula at nonzero temperature without recourse to
the fluctuation-dissipation theorem is based on the direct
summation of the free energies of all photon oscillator
modes \cite{GKM_03} (at zero temperature this method was
proposed in Refs.~\cite{rev106,rev107}; see also the
generalizations in Refs.~\cite{N_P_W,rev108,KMM_00}).

Equations for the determination of the frequencies of
oscillator modes between two dissimilar plates are obtained
from the Maxwell equations with the corresponding boundary
conditions on the boundary planes $z=\pm a/2$. They are given by
\cite{rev108,KMM_00}
\begin{eqnarray}
&&
\Delta_{\|}(\omega,k_{\bot})\equiv
\left[k^{(1)}+{\ee}^{(1)}(\omega)q\right]
\left[k^{(2)}+{\ee}^{(2)}(\omega)q\right]e^{aq}
\nonumber \\
&&\phantom{aa}
-
\left[k^{(1)}-{\ee}^{(1)}(\omega)q\right]
\left[k^{(2)}-{\ee}^{(2)}(\omega)q\right]e^{-aq}=0,
\nonumber \\
&&
\Delta_{\bot}(\omega,k_{\bot})\equiv
\left(k^{(1)}+q\right)
\left(k^{(2)}+q\right)e^{aq}
\nonumber \\
&&\phantom{aa}
-
\left(k^{(1)}-q\right)
\left(k^{(2)}-q\right)e^{-aq}=0,
\label{eqI8}
\end{eqnarray}
where
\begin{equation}
k^{(k)}=\sqrt{k_{\bot}^2-{\ee}^{(k)}(\omega)
\frac{\omega^2}{c^2}},\qquad
q=\sqrt{k_{\bot}^2-\frac{\omega^2}{c^2}}.
\label{eqI9}
\end{equation}
\noindent
The solutions of Eq.~(\ref{eqI8}) can be denoted as
$\omega_{k_{\bot},n}^{\|}$ and $\omega_{k_{\bot},n}^{\bot}$.
The free energy of one oscillator mode is given by
\begin{eqnarray}
&&
{\cal F}_{k_{\bot},n}^{\|,\bot}(a,T)=
\frac{\hbar\omega_{k_{\bot},n}^{\|,\bot}}{2}+
k_BT\ln\left(1-
e^{-\frac{\hbar\omega_{k_{\bot},n}^{\|,\bot}}{k_BT}}
\right)
\nonumber \\
&&\phantom{aa}
=k_BT\ln\left(
2\sinh\frac{\hbar\omega_{k_{\bot},n}^{\|,\bot}}{2k_BT}
\right).
\label{eqI10}
\end{eqnarray}

After summation over all quantum numbers, we obtain
\begin{eqnarray}
&&
{\cal F}(a,T)=\frac{k_BT}{2\pi}
\sum\limits_{n}\int_{0}^{\infty}k_{\bot}dk_{\bot}
\left[\ln\left(
2\sinh\frac{\hbar\omega_{k_{\bot},n}^{\|}}{2k_BT}
\right)\right.
\nonumber \\
&&\phantom{aa}
\left.+
\ln\left(
2\sinh\frac{\hbar\omega_{k_{\bot},n}^{\bot}}{2k_BT}
\right)\right].
\label{eqI11}
\end{eqnarray}

Eq.~(\ref{eqI11}) can be identically represented by the use
of the argument theorem in terms of the quantities
$\Delta_{\|}(\omega,k_{\bot})$ and
$\Delta_{\bot}(\omega,k_{\bot})$ from Eq.~(\ref{eqI8})
having their roots at the frequencies of the oscillator
modes. After transformation of the branch points
into poles by means of integration by parts and
calculation of the residues at all imaginary Matsubara
frequencies $i\xi_l$, we rewrite Eq.~(\ref{eqI11})
in the form (see Ref.~\cite{GKM_03} for details)
\begin{eqnarray}
&&
{\cal F}(a,T)=\frac{k_BT}{2\pi}
\sum\limits_{l=0}^{\infty}
\left(1-\frac{1}{2}\delta_{l0}\right)
\int_{0}^{\infty}k_{\bot}dk_{\bot}
\nonumber \\
&&\phantom{aa}
\times
\left[\ln\Delta_{\|}(i\xi_l,k_{\bot})+
\ln\Delta_{\bot}(i\xi_l,k_{\bot})\right].
\label{eqI12}
\end{eqnarray}
\noindent
Expression (\ref{eqI12}) is infinite. To remove the divergences,
we subtract from the right-hand side of Eq.~(\ref{eqI12}) the
free energy in the case of infinitely remote plates
($a\to\infty$). Then Eq.~(\ref{eqI6}) with the notation
(\ref{eqI7}) is reobtained.

Both quantum field theoretical derivations of the Lifshitz formula
discussed above are immediately applicable for dielectrics
described by real dielectric permittivities ${\ee}^{(k)}(\omega)$.
In this case the unitarity condition in the first derivation
is valid and the photon oscillator modes in the second
derivation are real quantities. Both derivations, however,
under certain conditions, can be generalized to the case
of media with dissipation \cite{BG_75,M_T}.
In so doing a medium under the influence of electromagnetic
oscillations can be represented as a set of oscillators
\begin{equation}
\frac{d^2x(t)}{dt^2}+\gamma\frac{dx(t)}{dt}+
\omega_0^2x(t)=f(t),
\label{eqI13}
\end{equation}
\noindent
where $\gamma$ is a damping parameter and
$f(t)=f_{\omega}\exp(-i\omega t)$ is the harmonically changing
external force. Eq.~(\ref{eqI13}) has the solution
$x(t)=x_{\omega}\exp(-i\omega t)$ where
$x_{\omega}=\chi(\omega)f_{\omega}$, and
\begin{equation}
\chi(\omega)=\frac{1}{\omega_0^2-i\gamma\omega-\omega^2}
\label{eqI14}
\end{equation}
\noindent
is the susceptibility of the system. The latter is connected
with the dielectric permittivity of a medium
\begin{equation}
{\ee}(\omega)=1+\frac{2g}{\pi}\chi(\omega),
\label{eqI15}
\end{equation}
\noindent
where $g$ is the oscillator strength.

The dielectric permittivity (\ref{eqI15}) is complex and
describes an absorption band (or bands if one considers several
oscillators with different parameters) of finite width
and amplitude at nonzero frequency. Note that in
thermal equilibrium net heat losses are absent on
average \cite{BG_75,Ginzburg}. One can  conclude that for such
media all processes of absorption are balanced by
the respective processes of emission, and the global unitarity
condition remains valid.

Regarding the second field theoretical approach described above,
the oscillator frequency,
determined from the equation $1/\chi(\omega)=0$,
becomes complex when $\gamma\neq 0$.
In this case the free energy is not given
by Eq.~(\ref{eqI11}) (which is already clear from the
complexity of the right-hand side of this equation).
For the complex dielectric permittivity given by
Eq.~(\ref{eqI15}) the correct expression for the free
energy is obtained from the auxiliary electrodynamic
problem and leads to Eqs.~(\ref{eqI12}) and (\ref{eqI6})
(see Ref.~\cite{BG_75} for details). The qualitative
reason for the validity of this statement is that the
free energy depends only on the values of the dielectric
permittivity along the imaginary frequency axis [i.e.,
only on ${\ee}(i\xi)$] which are always real.
Note that these considerations are not applicable to
real metals described by the Drude dielectric function
because in this case the proper frequency $\omega_0$
of the respective oscillator
in Eq.~(\ref{eqI13}) turns into zero and the
absorption band is shifted to zero frequency and
achieves an infinite amplitude. As a result,
dissipation leads to heating of a metal and the unitarity
condition is violated (see Refs.~\cite{GKM_01,GKM_03}).

\section{Casimir effect for two plates made of dissimilar
dilute dielectrics}

In this section we anticipate that the dielectric permittivities
of both plates ($k=1,\,2$) are constant and equal to
${\ee}^{(k)}=1+\eta_k$, where $\eta_k\ll 1$.
The assumption that the dielectric permittivity does not depend
on frequency implies that the separation distance $a$ is rather
large (in fact larger than the characteristic absorption
wavelength) and, consequently,
relativistic effects are essential \cite{L_P}.
Thus, in this case we are dealing with the Casimir effect.

For convenience in analytic and numerical calculations, we
introduce the dimensionless variables $\zeta$ and $y$
given by
\begin{equation}
\zeta_l=\frac{\xi_l}{\xi_c}=\frac{2a\xi_l}{c}=\tau l,
\quad
y=2q_la,
\label{eq1a}
\end{equation}
\noindent
where $\xi_c=c/(2a)$ is the so called characteristic
frequency, $\tau=4\pi k_BaT/(\hbar c)$,
 and $q_l$ was defined in Eq.~(\ref{eqI5}). Then
the Lifshitz formula (\ref{eqI6}) takes the form
\begin{eqnarray}
&&
{\cal F}(a,T)=\frac{\hbar c\tau}{32\pi^2 a^3}
\sum\limits_{l=0}^{\infty}
\left(1-\frac{1}{2}\delta_{l0}\right)
\int_{\zeta_l}^{\infty}ydy
\nonumber \\
&&\phantom{aa}
\times\left\{
\ln\left[1-r_{\|}^{(1)}(\zeta_l,y)
r_{\|}^{(2)}(\zeta_l,y)e^{-y}\right]\right.
\label{eq1} \\
&&\phantom{aa}
+\left.
\ln\left[1-r_{\bot}^{(1)}(\zeta_l,y)
r_{\bot}^{(2)}(\zeta_l,y)e^{-y}\right]\right\}.
\nonumber
\end{eqnarray}
\noindent
Here the reflection coefficients (\ref{eqI7}) are expressed in
terms of new variables (\ref{eq1a}) as follows:
\begin{eqnarray}
&&
r_{\|}^{(k)}(\zeta_l,y)=
\frac{{\ee}_l^{(k)}y-
\sqrt{y^2+\zeta_l^2({\ee}_l^{(k)}-1)}}{{\ee}_l^{(k)}y+
\sqrt{y^2+\zeta_l^2({\ee}_l^{(k)}-1)}}\,,
\nonumber \\
&&\phantom{aa}
r_{\bot}^{(k)}(\zeta_l,y)=
\frac{\sqrt{y^2+\zeta_l^2({\ee}_l^{(k)}-1)}-
y}{\sqrt{y^2+\zeta_l^2({\ee}_l^{(k)}-1)}+y}\,.
\label{eq2}
\end{eqnarray}
\noindent
Note that
${\ee}_l^{(k)}\equiv{\ee}^{(k)}(i\xi_l)=
{\ee}^{(k)}(i\zeta_l\xi_c)$.
In the case of constant dielectric permittivities,
which are under
consideration in this section,
${\ee}_l^{(k)}={\ee}^{(k)}$, i.e., being independent on $l$.

Substitution of ${\ee}^{(k)}=1+\eta_k$ in Eq.~(\ref{eq1}) and
expansion of the logarithms in powers of the small parameters
$\eta_1$, $\eta_2$ (preserving all powers up to order 3 inclusive)
leads to
\begin{eqnarray}
&&
{\cal F}(a,T)=-\frac{\hbar c\tau\eta_1\eta_2}{256\pi^2 a^3}
\sum\limits_{l=0}^{\infty}
\left(1-\frac{1}{2}\delta_{l0}\right)
\nonumber \\
&&\phantom{aa}
\times\left[I_1(\zeta_l)-\frac{\eta_1+\eta_2}{2}I_2(\zeta_l)
\right],
\label{eq3}
\end{eqnarray}
\noindent
where
\begin{eqnarray}
&&
I_1(\zeta_l)\equiv\int_{\zeta_l}^{\infty}dy
\frac{e^{-y}}{y^3}\left(2y^4-2\zeta_l^2y^2+\zeta_l^4\right)
\nonumber \\
&&\phantom{aa}
=e^{-\zeta_l}\left(2+2\zeta_l+\frac{\zeta_l^2}{2}-
\frac{\zeta_l^3}{2}\right)+\zeta_l^2\left(2-
\frac{\zeta_l^2}{2}\right)\mbox{Ei}(-\zeta_l),
\nonumber \\
&&
I_2(\zeta_l)\equiv\int_{\zeta_l}^{\infty}dy
\frac{e^{-y}}{y^5}\left(2y^6-\zeta_l^2y^4-\zeta_l^4y^2+\zeta_l^6\right)
\label{eq5} \\
&&\phantom{aa}
=e^{-\zeta_l}\left(2+2\zeta_l-\frac{\zeta_l^2}{4}+
\frac{5\zeta_l^3}{12}+\frac{\zeta_l^4}{24}-
\frac{\zeta_l^5}{24}\right)
\nonumber \\
&&\phantom{aa}
+\zeta_l^2\left(1+\frac{\zeta_l^2}{2}-
\frac{\zeta_l^4}{24}\right)\mbox{Ei}(-\zeta_l),
\nonumber
\end{eqnarray}
\noindent
and $\mbox{Ei}(z)$ is the exponential integral function.

Summation in Eq.~(\ref{eq3}) can be performed in the following
way:
\begin{eqnarray}
&&
f_1(\tau)\equiv
\sum\limits_{l=0}^{\infty}
\left(1-\frac{1}{2}\delta_{l0}\right)I_1(\zeta_l)
\label{eq6} \\
&&\phantom{aa}
=1+2\frac{(1+\tau)e^{\tau}-1}{(e^{\tau}-1)^2}
+\frac{\tau^2e^{\tau}\left[(1-\tau)e^{2\tau}-
4\tau e^{\tau}-\tau-1\right]}{2(e^{\tau}-1)^4}
\nonumber \\
&&\phantom{aa}
+\sum\limits_{l=1}^{\infty}F_1(\zeta_l),
\nonumber \\
&&
f_2(\tau)\equiv
\sum\limits_{l=0}^{\infty}
\left(1-\frac{1}{2}\delta_{l0}\right)I_2(\zeta_l)
\label{eq7} \\
&&\phantom{aa}
=1+2\frac{(1+\tau)e^{\tau}-1}{(e^{\tau}-1)^2}
+\frac{\tau^2e^{\tau}\left[(5\tau -3)e^{2\tau}+
20\tau e^{\tau}+5\tau+3\right]}{12(e^{\tau}-1)^4}
\nonumber \\
&&\phantom{aa}
-\frac{\tau^4e^{\tau}\left[(\tau -1)e^{4\tau}+
2(13\tau -5)e^{3\tau}+
66\tau e^{2\tau}+2(13\tau +5)e^{\tau}+
\tau+1\right]}{24(e^{\tau}-1)^6}
\nonumber \\
&&\phantom{aa}
+\sum\limits_{l=1}^{\infty}F_2(\zeta_l),
\nonumber
\end{eqnarray}
\noindent
where
\begin{eqnarray}
&&
F_1(\zeta_l)=\zeta_l^2\left(2-
\frac{\zeta_l^2}{2}\right)\mbox{Ei}(-\zeta_l),
\label{eq8} \\
&&
F_2(\zeta_l)=\zeta_l^2\left(1+\frac{\zeta_l^2}{2}-
\frac{\zeta_l^4}{24}\right)\mbox{Ei}(-\zeta_l).
\nonumber
\end{eqnarray}

Thus, the exact (in $\tau$) expression for the Casimir
free energy for the configuration of two dissimilar dilute
dielectric plates is given by
\begin{equation}
{\cal F}(a,T)=-\frac{\hbar c\tau\eta_1\eta_2}{256\pi^2 a^3}
\left[f_1(\tau)-\frac{\eta_1+\eta_2}{2}f_2(\tau)\right],
\label{eq8a}
\end{equation}
\noindent
where $f_{1,2}(\tau)$ are defined in Eqs.~(\ref{eq6})--(\ref{eq8}).

It is not difficult to find the asymptotic behavior of
Eq.~(\ref{eq8a}) at $\tau\ll 1$. For this purpose the
remaining sums on the right-hand side of Eqs.~(\ref{eq6}),
(\ref{eq7}) are calculated with the help of the
Abel-Plana formula \cite{M_T,B_M_M}
\begin{eqnarray}
&&
\sum\limits_{l=0}^{\infty}
\left(1-\frac{1}{2}\delta_{l0}\right)F(l)=
\int_{0}^{\infty}F(t)dt
\nonumber \\
&&\phantom{aa}
+i\int_{0}^{\infty}dt
\frac{F(it)-F(-it)}{e^{2\pi t}-1},
\label{eq9}
\end{eqnarray}
\noindent
where $F(z)$ is an analytic function in the right half-plane.
Taking into account that $F_{1,2}(0)=0$, these sums are
equal to
\begin{eqnarray}
&&
\sum\limits_{l=1}^{\infty}
F_{1,2}(\zeta_l)=
\int_{0}^{\infty}F_{1,2}(\tau t)dt
\nonumber \\
&&\phantom{aa}
+i\int_{0}^{\infty}dt
\frac{F_{1,2}(i\tau t)-F_{1,2}(-i\tau t)}{e^{2\pi t}-1}.
\label{eq10}
\end{eqnarray}
\noindent
Using Eq.~(\ref{eq8}) one obtains
\begin{equation}
\int_{0}^{\infty}F_{1}(\tau t)dt=\frac{16}{15\tau},
\quad
\int_{0}^{\infty}F_{2}(\tau t)dt=\frac{128}{105\tau}.
\label{eq11}
\end{equation}
\noindent
The second integral on the right-hand side of Eq.~(\ref{eq10})
can be calculated perturbatively. From Eq.~(\ref{eq8})
it follows:
\begin{eqnarray}
&&
F_{1}(i\tau t)-F_{1}(-i\tau t)=-2i\pi\tau^2t^2+
4i\tau^3t^3+O(\tau^4),
\label{eq12} \\
&&
F_{2}(i\tau t)-F_{2}(-i\tau t)=-i\pi\tau^2t^2+
2i\tau^3t^3+O(\tau^4).
\nonumber
\end{eqnarray}
\noindent
Substituting Eqs.~(\ref{eq11}) and (\ref{eq12}) into
Eq.~(\ref{eq10}), we arrive at
\begin{eqnarray}
&&
\sum\limits_{l=1}^{\infty}
F_{1}(\zeta_l)=\frac{16}{15\tau}+\frac{\zeta(3)}{2\pi^2}\tau^2
-\frac{\tau^3}{60}+O(\tau^4),
\label{eq13} \\
&&
\sum\limits_{l=1}^{\infty}
F_{2}(\zeta_l)=\frac{128}{105\tau}+\frac{\zeta(3)}{4\pi^2}\tau^2
-\frac{\tau^3}{120}+O(\tau^4),
\nonumber
\end{eqnarray}
\noindent
where $\zeta(z)$ is the Riemann zeta-function.
After expansion of the remaining terms on the right-hand
sides of Eqs.~(\ref{eq6}), (\ref{eq7}) in powers of $\tau$,
one obtains
\begin{eqnarray}
&&
f_{1}(\tau)=\frac{46}{15\tau}+\frac{\zeta(3)}{2\pi^2}\tau^2
-\frac{7\tau^3}{360}+O(\tau^4),
\label{eq14} \\
&&
f_{2}(\tau)=\frac{338}{105\tau}+\frac{\zeta(3)}{4\pi^2}\tau^2
+\frac{\tau^3}{360}+O(\tau^4).
\nonumber
\end{eqnarray}

The substitution of Eq.~(\ref{eq14}) in Eq.~(\ref{eq8a})
leads to the final result
\begin{eqnarray}
&&
{\cal F}(a,T)=-\frac{\hbar c\eta_1\eta_2}{256\pi^2 a^3}
\left[\frac{46}{15}+\frac{\zeta(3)}{2\pi^2}\tau^3
-\frac{7\tau^4}{360}\right.
\label{eq15} \\
&&\phantom{aa}\left.
-\frac{\eta_1+\eta_2}{2}\left(
\frac{338}{105}+\frac{\zeta(3)}{4\pi^2}\tau^3
+\frac{\tau^4}{360}\right)+O(\tau^5)\right].
\nonumber
\end{eqnarray}

The asymptotic behavior of the Casimir free energy at
$\tau\gg 1$ (high temperatures or large separations)
also can be obtained from Eq.~(\ref{eq8a}).
With the proviso that $\tau\gg 1$, Eqs.~(\ref{eq6})
and (\ref{eq7}) lead to $f_1(\tau)=f_2(\tau)=1$
up to exponentially small corrections. As a result
\begin{equation}
{\cal F}(a,T)=-\frac{k_BT\eta_1\eta_2}{64\pi a^2}
\left(1-\frac{\eta_1+\eta_2}{2}\right).
\label{eq16}
\end{equation}

The same expression follows   from the zero-frequency
contribution to the free energy in Eq.~(\ref{eq1})
\begin{equation}
{\cal F}(a,T)=\frac{\hbar c\tau}{64\pi a^3}
\int_{0}^{\infty}ydy\ln
\left[1-\frac{\eta_1\eta_2}{(2+\eta_1)(2+\eta_2)}\,e^{-y}\right]
\label{eq17}
\end{equation}
\noindent
after integration with respect to $y$ and expansion
in powers of $\eta_1$ and $\eta_2$.
Note that at zero frequency only $r_{\|}^{(k)}$ contributes
to Eq.~(\ref{eq17}), because, according to Eq.~(\ref{eq2}),
$r_{\bot}^{(k)}(0,y)=0$.

Now let us consider the Casimir pressure and entropy for the
configuration of two dissimilar dielectric plates.
The Lifshitz formula for the Casimir pressure presented in
terms of dimensionless variables (\ref{eq1a}) has the form
\begin{eqnarray}
&&
P(a,T)=-\frac{\hbar c\tau}{32\pi^2a^4}
\sum\limits_{l=0}^{\infty}
\left(1-\frac{1}{2}\delta_{l0}\right)
\int_{\zeta_l}^{\infty}y^2dy
\nonumber \\
&&\phantom{aa}
\times\left[\frac{r_{\|}^{(1)}(\zeta_l,y)
r_{\|}^{(2)}(\zeta_l,y)}{e^{y}-r_{\|}^{(1)}(\zeta_l,y)
r_{\|}^{(2)}(\zeta_l,y)}+
\frac{r_{\bot}^{(1)}(\zeta_l,y)
r_{\bot}^{(2)}(\zeta_l,y)}{e^{y}-r_{\bot}^{(1)}(\zeta_l,y)
r_{\bot}^{(2)}(\zeta_l,y)}\right].
\label{eq18}
\end{eqnarray}

Expanding in powers of
small parameters $\eta_1$, $\eta_2$ by the same
procedure as in the case of the free energy, we obtain the
expression exact in $\tau$
\begin{equation}
{P}(a,T)=-\frac{\hbar c\tau\eta_1\eta_2}{256\pi^2 a^4}
\left[p_1(\tau)-\frac{\eta_1+\eta_2}{2}p_2(\tau)\right],
\label{eq19}
\end{equation}
\noindent
where $p_{1,2}(\tau)$ are given by
\begin{eqnarray}
&&
p_1(\tau)
=2+4\frac{(1+\tau)e^{\tau}-1}{(e^{\tau}-1)^2}
+\frac{\tau^3e^{\tau}\left(e^{2\tau}+
4e^{\tau}+1\right)}{(e^{\tau}-1)^4}
\nonumber \\
&&\phantom{aa}
+\tau^4\sum\limits_{l=1}^{\infty}l^4\mbox{Ei}(-\tau l),
\label{eq20} \\
&&
p_2(\tau)
=2+4\frac{(1+\tau)e^{\tau}-1}{(e^{\tau}-1)^2}
-\frac{\tau^2e^{\tau}\left[(2\tau -3)e^{2\tau}+
8\tau e^{\tau}+2\tau+3\right]}{3(e^{\tau}-1)^4}
\nonumber \\
&&\phantom{aa}
+\frac{\tau^4e^{\tau}\left[(\tau -1)e^{4\tau}+
2(13\tau -5)e^{3\tau}+
66\tau e^{2\tau}+2(13\tau +5)e^{\tau}+
\tau+1\right]}{6(e^{\tau}-1)^6}
\nonumber \\
&&\phantom{aa}
-\tau^4\sum\limits_{l=1}^{\infty}l^4\left(1-
\frac{\tau^2l^2}{6}\right)\mbox{Ei}(-\tau l).
\nonumber
\end{eqnarray}

In the limiting case $\tau\ll 1$, Eqs.~(\ref{eq19}), (\ref{eq20})
lead to
\begin{eqnarray}
&&
{P}(a,T)=-\frac{\hbar c\eta_1\eta_2}{256\pi^2 a^4}
\left[\frac{46}{5}+\frac{7\tau^4}{360}\right.
\label{eq21} \\
&&\phantom{aa}\left.
-\frac{\eta_1+\eta_2}{2}\left(
\frac{338}{35}-\frac{\tau^4}{360}\right)+O(\tau^5)\right].
\nonumber
\end{eqnarray}
\noindent
This expression can also  be obtained as
\begin{equation}
P(a,T)=-\frac{\partial{\cal F}(a,T)}{\partial a},
\label{eq22}
\end{equation}
\noindent
where ${\cal F}(a,T)$ is given by Eq.~(\ref{eq15}).

At high temperatures (large separations) it holds $\tau\gg 1$
and from Eqs.~(\ref{eq19}), (\ref{eq20}) it follows:
\begin{equation}
P(a,T)=-\frac{k_BT\eta_1\eta_2}{32\pi a^3}
\left(1-\frac{\eta_1+\eta_2}{2}\right).
\label{eq23}
\end{equation}

Using the above procedure, one can obtain the Casimir
entropy for the configuration of two
dissimilar dilute dielectric plates.
It can be found also as
\begin{equation}
S(a,T)=-\frac{\partial{\cal F}(a,T)}{\partial T},
\label{eq24}
\end{equation}
\noindent
where ${\cal F}(a,T)$ is given by Eq.~(\ref{eq8a}).
In the limit $\tau\ll 1$ the result is
\begin{eqnarray}
&&
S(a,T)=\frac{3k_B\zeta(3)\eta_1\eta_2\tau^2}{128\pi^3a^2}
\left[
\vphantom{\left(\frac{\eta_1+\eta_2}{14}\right)
\frac{7\pi^2}{135\zeta(3)}}
1-\frac{\eta_1+\eta_2}{4}\right.
\label{eq25} \\
&&\phantom{aa}\left.
-\left(1+\frac{\eta_1+\eta_2}{14}\right)
\frac{7\pi^2\tau}{135\zeta(3)}+O(\tau^2)\right].
\nonumber
\end{eqnarray}
\noindent
As is seen from Eq.~(\ref{eq25}), $S(a,0)=0$, i.e.,
the Nernst heat theorem is satisfied.

The obtained results (\ref{eq15}), (\ref{eq21}), and
(\ref{eq25}) are used as the tests in the next section
where the low-temperature behavior the Casimir and
van der Waals interactions
between plates with arbitrary (not dilute) dielectric
permittivities is studied.

\section{Low-temperature behavior of the van der Waals
and Casimir interactions between dissimilar dielectrics:
analytical results}

We will now look at the configuration of two parallel plates
made of real dissimilar dielectrics described by
frequency-dependent dielectric permittivities. A distinguishing
feature of dielectrics is the finite value of their static
dielectric permittivity ${\ee}_0^{(k)}={\ee}^{(k)}(0)$.
It is common for dielectrics that the dielectric permittivity is
practically equal to its static value within some frequency
region $[0,\omega_k]$. At higher frequencies
${\ee}^{(k)}(\omega)$ is smaller than
${\ee}_0^{(k)}$.

We start with the Lifshitz formula (\ref{eq1}), (\ref{eq2})
which is applicable to any dielectric plates at any separation
(note that separations $a$ should be large enough so
one can ignore the atomic structure of the plates,
i.e., $a$ should be larger than 2 or 3\,nm).

Applying the Abel-Plana formula (\ref{eq9}), we can identically
rearrange Eq.~(\ref{eq1}) to the form
\begin{equation}
{\cal F}(a,T)=E(a)+\Delta{\cal F}(a,T).
\label{eq26}
\end{equation}
\noindent
Here $E(a)$ is the energy of the van der Waals or Casimir
interaction at zero temperature given by
\begin{equation}
E(a)=\frac{\hbar c}{32\pi^2a^3}
\int_{0}^{\infty}d\zeta\int_{\zeta}^{\infty}
dy\,f(\zeta,y),
\label{eq27}
\end{equation}
\noindent
where we introduce the notation
\begin{eqnarray}
&&
f(\zeta,y)=y\left\{\ln\left[1-r_{\|}^{(1)}(\zeta,y)
r_{\|}^{(2)}(\zeta,y)e^{-y}\right]\right.
\nonumber \\
&&\phantom{aa}\left.
+\ln\left[1-r_{\bot}^{(1)}(\zeta,y)
r_{\bot}^{(2)}(\zeta,y)e^{-y}\right]\right\}.
\label{eq28}
\end{eqnarray}
\noindent
The second contribution on the right-hand side in
Eq.~(\ref{eq26}) is the thermal correction to the energy
\begin{equation}
\Delta{\cal F}(a,T)=\frac{i\hbar c\tau}{32\pi^2a^3}
\int_{0}^{\infty}dt\frac{F(i\tau t)-
F(-i\tau t)}{e^{2\pi t}-1},
\label{eq29}
\end{equation}
\noindent
where the function $F(x)$ is defined by
\begin{equation}
F(x)=\int_{x}^{\infty}dy\,f(x,y).
\label{eq30}
\end{equation}

The behavior of the energy (\ref{eq27}) at both short
separations (the limit of the van der Waals forces) and
large separations (the limit of the Casimir forces) is
well-studied \cite{L_P}. At short separations
($a\ll\lambda_0$) the dependence of the dielectric
permittivities on frequency is important and the asymptotic
behavior of the energy is given by
$E(a)=-H/(12\pi a^2)$, where $H$ is the Hamaker constant
\cite{L_P,B_M_M}. At large separations
($a\gg\lambda_0$) ${\ee}^{(k)}(\omega)$ can be approximated
by their
static values ${\ee}_0^{(k)}$. As a result, the energy
takes the form $E(a)=-\Psi(\ezf,\ezs)/a^3$,
where $\Psi$ is some tabulated function of two
variables \cite{L_P,B_M_M}.

Importantly, the low-temperature behavior of the thermal
correction (\ref{eq29}) under the condition $\tau\ll 1$ has a
universal form valid at both short and large separations.
The qualitative explanation for this fact, proven below in
detail, is that for dielectrics at sufficiently low temperatures
the Matsubara frequencies contributing to the thermal
correction belong to the region where ${\ee}^{(k)}$
practically do not depend on the frequency and are equal to
their static values ${\ee}_0^{(k)}$.

To prove the universal low-temperature character of the
thermal correction, we use the Ninham-Parsegian representation
for the dielectric permittivities along the imaginary
frequency axis \cite{N_P,M_N,Berg},
\begin{equation}
{\ee}^{(k)}(i\xi)=1+\sum\limits_{j}
\frac{C_j^{(k)}}{1+\frac{\xi^2}{{\omega_j^{(k)}}^2}}.
\label{eq31}
\end{equation}
\noindent
Here, the parameters $C_j^{(k)}$ are the absorption strengths for
different dielectrics ($k=1,\,2$) satisfying the condition,
\begin{equation}
\sum\limits_{j}
{C_j^{(k)}}={\ee}_0^{(k)}-1,
\label{eq31a}
\end{equation}
\noindent
and $\omega_j^{(k)}$ are the characteristic absorption frequencies.
Although Eq.~(\ref{eq31}) is the approximate one, it
gives a very accurate description for many dielectrics \cite{Berg}.

To obtain the asymptotic behavior of the thermal correction to
the energy given by Eq.~(\ref{eq29}) at $\tau\ll 1$, we
substitute Eq.~(\ref{eq31}) in Eq.~(\ref{eq28}) and expand the
function $f(x,y)$ in powers of $x=\tau t$. The subsequent
integration of this expansion with respect to $y$ from $x$ to
infinity (see Appendix A) leads to
\begin{eqnarray}
&&
F(ix)-F(-ix)=i\pi
\frac{\ezf +\ezs +2\ezf \ezs}{(\ezf +1)(\ezs +1)}\,
\frac{(\ezf -1)(\ezs -1)}{2(\ezf +\ezs)}\,x^2
\nonumber \\
&&\phantom{aa}
-i\alpha x^3+O(x^4).
\label{eq32}
\end{eqnarray}
\noindent
On the right-hand side of Eq.~(\ref{eq32}) we separated the
third-order contribution with a real coefficient $\alpha$.
It cannot be determined at this stage of our calculations
because all terms in the expansion of $f(x,y)$ in powers
of $x$ contribute to this coefficient [the value of $\alpha$
is found below in Eq.~(\ref{eq44})].
As is seen from Eq.~(\ref{eq32}), only the static dielectric
permittivities contribute to $F(ix)-F(-ix)$ in leading
order and, consequently, to the thermal correction.

Substituting Eq.~(\ref{eq32}) in Eqs.~(\ref{eq29}) and (\ref{eq26})
 and performing the integration, we arrive at the result
\begin{eqnarray}
&&
{\cal F}(a,T)=E(a)-\frac{\hbar c}{32\pi^2a^3}
\label{eq33} \\
&&\phantom{aa}\times
\left[\frac{\zeta(3)\left(\ezf +\ezs +
2\ezf \ezs\right)}{(\ezf +1)(\ezs +1)}\,
\frac{(\ezf -1)(\ezs -1)}{8\pi^2(\ezf +\ezs)}\,\tau^3
-C_4\tau^4+O(\tau^5)\right],
\nonumber
\end{eqnarray}
\noindent
where $C_4\equiv\alpha/240$.

So far we have considered the free energy. The pressure can be
obtained, using Eqs.~(\ref{eq22}) and (\ref{eq33}), as
\begin{equation}
P(a,T)=P_0(a)-\frac{\hbar c}{32\pi^2a^4}\left[
C_4\tau^4+O(\tau^5)\right],
\label{eq34}
\end{equation}
\noindent
where $P_0(a)=-\partial E(a)/\partial a$ is the pressure
at zero temperature.
Our aim is to determine the value of $C_4$. To attain this,
the pressure is expressed directly by the Lifshitz
formula (\ref{eq18}). Applying the Abel-Plana formula
(\ref{eq9}) in Eq.~(\ref{eq18}), we represent the pressure
in the form
\begin{equation}
P(a,T)=P_0(a)+\Delta P(a,T),
\label{eq35}
\end{equation}
\noindent
where the thermal correction to the pressure is
\begin{equation}
\Delta P(a,T)=-\frac{i\hbar c\tau}{32\pi^2a^4}
\int_{0}^{\infty}dt\frac{\Phi(i\tau t)-
\Phi(-i\tau t)}{e^{2\pi t}-1}
\label{eq36}
\end{equation}
\noindent
and the function $\Phi(x)\equiv \Phi_{\|}(x)+\Phi_{\bot}(x)$
is given by
\begin{equation}
\Phi_{\|,\bot}(x)=\int_{x}^{\infty}
\frac{y^2dy\,r_{\|,\bot}^{(1)}(x,y)
r_{\|,\bot}^{(2)}(x,y)}{e^{y}-r_{\|,\bot}^{(1)}(x,y)
r_{\|,\bot}^{(2)}(x,y)}\, .
\label{eq37}
\end{equation}

First, let us determine the leading term of the expansion of
$\Phi_{\bot}(x)$ in powers of $x$. By introducing the new
variable $v=y/x$ one arrives at
\begin{equation}
\Phi_{\bot}(x)=x^3\int_{1}^{\infty}
\frac{v^2dv\,r_{\bot}^{(1)}(x,v)
r_{\bot}^{(2)}(x,v)}{e^{vx}-r_{\bot}^{(1)}(x,v)
r_{\bot}^{(2)}(x,v)}\, .
\label{eq38}
\end{equation}
\noindent
Note that the reflection coefficients $r_{\bot}^{(k)}(x,v)$ depend
on $x$ only through the frequency dependence of ${\ee}^{(k)}$
according to Eq.~(\ref{eq31}). Expanding in powers of $x$ in
Eq.~(\ref{eq38}), we obtain
\begin{equation}
\Phi_{\bot}(x)=x^3\int_{1}^{\infty}
\frac{v^2dv\,r_{\bot}^{(1)}(0,v)
r_{\bot}^{(2)}(0,v)}{1-r_{\bot}^{(1)}(0,v)
r_{\bot}^{(2)}(0,v)}\,+O(x^4),
\label{eq39}
\end{equation}
\noindent
where, according to Eq.~(\ref{eq2}),
\begin{equation}
r_{\bot}^{(k)}(v,0)=\frac{\sqrt{v^2+{\ee}_0^{(k)} -1}-
v}{\sqrt{v^2+{\ee}_0^{(k)} -1}+v}.
\label{eq40}
\end{equation}
\noindent
Integration of Eq.~(\ref{eq39}) with account of Eq.~(\ref{eq40})
results in
\begin{equation}
\Phi_{\bot}(x)=\left(1-
\frac{\ezf +\ezs +\sqrt{\ezf \ezs}-\ezf \ezs}{\sqrt{\ezf}+
\sqrt{\ezs}}\right)\,\frac{x^3}{6}
+O(x^4).
\label{eq41}
\end{equation}
\noindent
From  Eq.~(\ref{eq41}) it follows
\begin{equation}
\Phi_{\bot}(ix)-\Phi_{\bot}(-ix)=-i\left(1-
\frac{\ezf +\ezs +\sqrt{\ezf \ezs}-\ezf \ezs}{\sqrt{\ezf}+
\sqrt{\ezs}}\right)\,\frac{x^3}{3}
+O(x^5).
\label{eq42}
\end{equation}

The expansion of $\Phi_{\|}(x)$ from Eq.~(\ref{eq37}) in
powers of $x$ is somewhat more cumbersome. It is presented
in detail in Appendix B leading to the result
\begin{eqnarray}
&&
\Phi_{\|}(ix)-\Phi_{\|}(-ix)=-i\left\{1+
\frac{1}{\left(\sqrt{\ezf}+
\sqrt{\ezs}\right)\left(\ezf +\ezs \right)^2}
\left[
\vphantom{\frac{{\ezf}^2 {\ezs}^2\left(\ezf -1\right)\left(\ezs -
1\right)}{\left(\sqrt{\ezf} -\sqrt{\ezs}\right)\sqrt{\ezf +\ezs}}}
-\left({\ezf +\ezs}\right)^3
\right.\right.
\label{eq43} \\
&&\phantom{a}
+\ezf \ezs \sqrt{\ezf \ezs}
\left(5\ezf \ezs -3\ezf -3\ezs+1\right)
\nonumber \\
&&\phantom{aa}
+\sqrt{\ezf \ezs}\left(\sqrt{\ezf}-\sqrt{\ezs}\right)^2
\left(\ezf \ezs \sqrt{\ezf \ezs}-\sqrt{\ezf \ezs}-\ezf
-\ezs\right)
\nonumber \\
&&\phantom{aa}
\left.\left.
-
\frac{3{\ezf}^2 {\ezs}^2\left(\ezf -1\right)\left(\ezs -
1\right)}{\left(\sqrt{\ezf} -\sqrt{\ezs}\right)\sqrt{\ezf +\ezs}}
\mbox{Artanh}\frac{\sqrt{\ezf +\ezs}\left(\sqrt{\ezf} -
\sqrt{\ezs}\right)}{\sqrt{\ezf \ezs}-\ezf -\ezs}\right]\right\}\,
\frac{x^3}{3}+O(x^4).
\nonumber
\end{eqnarray}

Adding Eqs.~(\ref{eq42}) and (\ref{eq43}) and integrating the
obtained result according to Eq.~(\ref{eq36}), we find the
coefficient $C_4$ in the thermal correction to the energy
(\ref{eq33}) and pressure (\ref{eq34})
\begin{eqnarray}
&&
C_4=\frac{1}{720}
\left\{2+
\frac{1}{\left(\sqrt{\ezf}+
\sqrt{\ezs}\right)\left(\ezf +\ezs \right)^2}
\left[
\vphantom{\frac{{\ezf}^2 {\ezs}^2\left(\ezf -1\right)\left(\ezs -
1\right)}{\left(\sqrt{\ezf} -\sqrt{\ezs}\right)\sqrt{\ezf +\ezs}}}
-\left({\ezf +\ezs}\right)^2
\right.\right.
\label{eq44} \\
&&\phantom{a}
\times\left(2\ezf +2\ezs +\sqrt{\ezf \ezs}-\ezf \ezs\right)
+\ezf \ezs \sqrt{\ezf \ezs}
\left(5\ezf \ezs -3\ezf -3\ezs+1\right)
\nonumber \\
&&\phantom{aa}
+\sqrt{\ezf \ezs}\left(\sqrt{\ezf}-\sqrt{\ezs}\right)^2
\left(\ezf \ezs \sqrt{\ezf \ezs}-\sqrt{\ezf \ezs}-\ezf
-\ezs\right)
\nonumber \\
&&\phantom{aa}
\left.\left.
-
\frac{3{\ezf}^2 {\ezs}^2\left(\ezf -1\right)\left(\ezs -
1\right)}{\left(\sqrt{\ezf} -\sqrt{\ezs}\right)\sqrt{\ezf +\ezs}}
\mbox{Artanh}\frac{\sqrt{\ezf +\ezs}\left(\sqrt{\ezf} -
\sqrt{\ezs}\right)}{\sqrt{\ezf \ezs}-\ezf -\ezs}\right]\right\}\, .
\nonumber
\end{eqnarray}

This expression gets much more simplified for the
case of two plates made of
one and the same dielectric ($\ezf =\ezs\equiv{\ee}_0$):
\begin{equation}
C_4=\frac{1}{720}\left(\sqrt{{\ee}_0}-1\right)
\left({\ee}_0^2+{\ee}_0\sqrt{{\ee}_0}-2\right).
\label{eq45}
\end{equation}
\noindent
Thus, for two similar dielectrics the free energy takes
the form
\begin{eqnarray}
&&
{\cal F}(a,T)=E(a)-\frac{\hbar c}{32\pi^2a^3}
\left[\frac{\zeta(3)({\ee}_0-1)^2}{8\pi^2({\ee}_0+1)}\tau^3
\right.
\label{eq46} \\
&&\phantom{aa}
\left.-\frac{1}{720}\left(\sqrt{{\ee}_0}-1\right)
\left({\ee}_0^2+{\ee}_0\sqrt{{\ee}_0}-2\right)\tau^4
+O(\tau^5)
\vphantom{\frac{\zeta(3)({\ee}_0-1)^2}{8\pi^2({\ee}_0+1)}}
\right].
\nonumber
\end{eqnarray}

Another limiting case is given by dilute dielectrics.
Putting ${\ee}_0^{(k)}=1+\eta_k$ in Eqs.~(\ref{eq33}),
(\ref{eq34}) and (\ref{eq44}) and expanding in powers of
the small parameters $\eta_1$, $\eta_2$, one arrives at
precisely the same thermal corrections (terms proportional
to $\tau^3$ and $\tau^4$) as were obtained for dilute
dielectrics in  Eqs.~(\ref{eq15}) and (\ref{eq21}).
Note that Eqs.~(\ref{eq15}) and (\ref{eq21}) also contain
terms independent of $\tau$ having a physical
meaning of the energy
and pressure at zero temperature. These terms coincide
with $E(a)$ and $P_0(a)$ from Eqs.~(\ref{eq33}) and
(\ref{eq34}) only in the relativistic limit of large
separations. This is because Eqs.~(\ref{eq15}) and
(\ref{eq21}) were obtained under the condition that the
dielectric permittivities are constant. However, as
indicated above, the thermal corrections obtained in
such a way have a universal character and are valid at
any separation with the constraint $\tau\ll 1$.

Eqs.~(\ref{eq33}), (\ref{eq44}) and (\ref{eq46}) solve
the vital issue about the thermodynamic consistency of the
Lifshitz formula for the case of two dielectric plates.
Using Eqs.~(\ref{eq33}) and (\ref{eq24}), the entropy of
the interaction between plates takes the form
\begin{eqnarray}
&&
S(a,T)=\frac{3k_B\zeta(3)(\ezf -1)(\ezs -
1)}{64\pi^3a^2(\ezf +1)}\,\tau^2\left[
\frac{\ezf +\ezs +2\ezf \ezs}{(\ezs +1)(\ezf +\ezs)}
\right.
\nonumber \\
&&\phantom{aa}
\left.
-\frac{32\pi^2(\ezf +1)C_4}{3\zeta(3)(\ezf -1)(\ezs -
1)}\,\tau+O(\tau^2)\right],
\label{eq47}
\end{eqnarray}
\noindent
where the coefficient $C_4$ is determined in Eq.~(\ref{eq44}).
As is seen from Eq.~(\ref{eq47}), the entropy vanishes when the
temperature goes to zero as it must be in accordance with
the third law of thermodynamics (the Nernst heat theorem).

In the limiting case of two similar dielectrics
Eq.~(\ref{eq47}) can be rearranged as follows:
\begin{eqnarray}
&&
S(a,T)=\frac{3k_B\zeta(3)({\ee}_0 -1)^2}{64\pi^3a^2({\ee}_0 +1)}
\,\tau^2
\nonumber \\
&&\phantom{aa}
\left[1
-\frac{2\pi^2({\ee}_0 +1)({\ee}_0\sqrt{{\ee}_0}+
2{\ee}_0+2\sqrt{{\ee}_0}+2)}{135\zeta(3)(\sqrt{{\ee}_0} +1)^2}\,
\tau+O(\tau^2)\right].
\label{eq48}
\end{eqnarray}
\noindent
In the limiting case of two dilute dielectrics the expansion
of Eq.~(\ref{eq47}) in powers of $\eta_1$ and $\eta_2$ coincides
with Eq.~(\ref{eq25}).

As is seen from Eqs.~(\ref{eq47}) and (\ref{eq48}), in the limit
$\tau\to 0$ ($T\to 0$) the lower order contributions to the
entropy are of the second and third powers in the small parameter
$\tau$. In this manner for dielectrics at low temperatures the
entropy obeys the same universal law which was previously found
for ideal \cite{Maclay,I2,Robash,R1} and real \cite{R1,R2} metals.
Recall that ideal metals (i.e., two plates with the Dirichlet
boundary conditions {\it{\bf E}}${}_t=0$) lead to an
exactly solvable model in Matsubara quantum field theory
\cite{Maclay,Mehra}. As was shown in Refs.~\cite{Robash,R1},
in the case of plates made of ideal metal the entropy at low
temperatures is given by
\begin{equation}
S(a,T)=\frac{3k_B\zeta(3)}{32\pi^3a^2}\tau^2
\left[1-\frac{2\pi^2}{135\zeta(3)}\tau+O(\tau^2)\right].
\label{eq48a}
\end{equation}
\noindent
This demonstrates that the thermal quantum field theory approach
is in perfect agreement with thermodynamics for ideal metals.
It should be pointed out that the expansion coefficients
of the free energy, pressure and entropy in powers of $\tau$
in the case of ideal metals cannot be obtained as a straightforward
limit ${\ee}_0^{(k)}\to\infty$ in Eqs.~(\ref{eq33}), (\ref{eq34})
and (\ref{eq47}). The mathematical reason is that it is
impermissible to interchange the limits $\tau\to 0$ and
${\ee}\to\infty$ in the power expansion of functions depending on
$\ee$ as a parameter.

In the case of large separations (high temperatures)
$\tau\gg 1$ and the approximation of static dielectric
permittivities is applicable. With this condition the major
contribution is given by the zero-frequency term of the Lifshitz
formula (\ref{eq1})
\begin{equation}
{\cal F}(a,T)=\frac{\hbar c\tau}{64\pi^2a^3}
\int_{0}^{\infty}ydy\ln\left[1-
\frac{\ezf -1}{\ezf +1}\,\frac{\ezs -1}{\ezs +1}\,
e^{-y}\right].
\label{eq49}
\end{equation}
\noindent
(the other terms being exponentially small). Integration in
Eq.~(\ref{eq49}) leads to
\begin{equation}
{\cal F}(a,T)=-\frac{k_BT}{16\pi a^2}
\mbox{Li}_3\left(
\frac{\ezf -1}{\ezf +1}\,\frac{\ezs -1}{\ezs +1}\right),
\label{eq50}
\end{equation}
\noindent
where $\mbox{Li}_n(z)$ is the polylogarithm function.
In a similar manner at $\tau\gg 1$ the pressure is given by
\begin{equation}
{P}(a,T)=-\frac{k_BT}{8\pi a^3}
\mbox{Li}_3\left(
\frac{\ezf -1}{\ezf +1}\,\frac{\ezs -1}{\ezs +1}\right).
\label{eq51}
\end{equation}
\noindent
In the limiting case of dilute dielectrics Eqs.~(\ref{eq50})
and (\ref{eq51}) coincide with Eqs.~(\ref{eq16})
and (\ref{eq23}), respectively.

\section{Comparison between analytic and numerical
results}

In this section we compare our analytic results in
Eqs.~(\ref{eq33})
and (\ref{eq34}) for the thermal corrections to the free energy
and pressure with computations performed for both dissimilar
and similar dielectrics using the complete tabulated optical
data and the Lifshitz formulas (\ref{eq1}), (\ref{eq18}).
This enables us to illustrate the applicability regions
of the above asymptotic expressions for different plate
materials and to gain an impression on the temperature
dependence of the free energy and pressure within a wide
range of parameters. As an example we consider two dielectrics
(Si and vitreous SiO${}_2$) which differ in their dielectric
properties.

The dielectric permittivity of real dielectrics along the
imaginary frequency axis can be obtained through the
dispersion relation
\begin{equation}
{\ee}(i\xi)=1+\frac{2}{\pi}
\int_{0}^{\infty}d\omega
\frac{\omega\,\mbox{Im}{\ee}(\omega)}{\omega^2+\xi^2}.
\label{eq52}
\end{equation}
Here $\mbox{Im}\,{\ee}(\omega)=2n_1(\omega)n_2(\omega)$,
where $n_1(\omega)=\mbox{Re}\,n(\omega)$,
$n_2(\omega)=\mbox{Im}\,n(\omega)$, and $n(\omega)$ is the complex
refractive index tabulated in Ref.~\cite{Palik}.
For the dielectric Si (of high resistivity
$\rho_0=1000\,\Omega\,$cm) and  vitreous SiO${}_2$ the
dielectric permittivities were computed by the use of
Eq.~(\ref{eq52}) in Ref.~\cite{C_K_M_Z}. The obtained results
are presented in Fig.~1 for Si (line 1) and for SiO${}_2$
(line 2). The flat steps in both lines should be extended for
all frequencies $0\leq\xi\leq 10^{10}\,$rad/s (the lowest
frequency indicated in Fig.~1). As a result, the static
dielectric permittivities are equal to $\ezf =11.66$ (for Si)
and $\ezs =3.84$ (for SiO${}_2$).

In Fig.~2, we compare the computational results for the
thermal corrections to the energy
$\Delta{\cal F}(a,T)={\cal F}(a,T)-E(a)$ (a) and pressure
$\Delta{P}(a,T)={P}(a,T)-P_0(a)$ (b) in the configuration of two
plates one made of Si and another one of SiO${}_2$ at
separation $a=400\,$nm as a function of the temperature.
Short-dashed lines are obtained by the use of the Lifshitz
formulas (\ref{eq1}),  (\ref{eq18}), (\ref{eq27}) with
the static dielectric permittivities ${\ee}_0^{(k)}$.
Solid lines are computed using the same Lifshitz formulas
with the complete frequency-dependent dielectric permittivities
of Si and SiO${}_2$ presented in Fig.~1. Long-dashed lines
show our asymptotic expressions for $\Delta{\cal F}(a,T)$
and $\Delta{P}(a,T)$ on the right-hand sides of
Eqs.~(\ref{eq33}) and (\ref{eq34}), respectively, with a
coefficient $C_4$ defined in Eq.~(\ref{eq44}).

 As is seen from Fig.~2a, the thermal correction to the
energy can be calculated with the static dielectric
permittivities of Si and SiO${}_2$ at temperatures below
100\,K. At higher temperatures the solid line computed
by the use of the tabulated optical data departs from the
short-dashed line computed with the static dielectric
permittivities. At $T<60\,$K the asymptotic expression for
$\Delta{\cal F}(a,T)$ in Eq.~(\ref{eq33}) with the
coefficient (\ref{eq44}) leads to the same results as the
original Lifshitz formula. Quite similar situation takes
place for the thermal correction to the pressure (see
Fig.~2b). The numerical results obtained by the use of
the Lifshitz formula with static ${\ee}^{(k)}$ coincide
with those obtained using the tabulated optical data
at $T<100\,$K (solid and short-dashed lines).
As to the asymptotic expression  for the thermal
correction to the pressure [Eqs.~(\ref{eq34}) and
(\ref{eq44})], it becomes applicable at $T<50\,$K.

To give a more comprehensive idea on the application regions
of the obtained asymptotic expressions, in Fig.~3 we present
the same information, as in Fig.~2, for the case of two similar
plates made of vitreous SiO${}_2$ at separation
$a=450\,$nm. As before, in Fig.~3a the thermal correction
to the energy is plotted and in Fig.~3b to the pressure
versus temperature. As is seen from Fig.~3a, all three
theoretical descriptions of the free energy (in terms of
the tabulated data, static dielectric permittivity and
the asymptotic one) become applicable at $T<85\,$K.
If we consider the thermal correction to the pressure
(Fig.~3b), different models of $\ee$ are applicable
at $T<100\,$K whereas the asymptotic expression
(\ref{eq34}) with the coefficient (\ref{eq45}) coincides
with them at $T<65\,$K. The distinguishing feature of
Fig.~3b is the intersection of the solid and long-dashed
lines. It is explained by a greater deviation of the solid
line, computed by the use of tabulated data, from the
short-dashed line, computed using static $\ezs$, because
of the presence of a second flat step in the frequency
dependence of ${\ee}^{(2)}(i\xi)$ (see line 2 in Fig.~1).

In Fig.~4 we present the dependences of the thermal
correction to the energy (a) and pressure (b) for two Si
plates at separation $a=300\,$nm computed in the
framework of the same approaches. For two Si plates
within the separation and temperature region under
consideration the lines computed with frequency-dependent
and static ${\ee}^{(1)}$ almost coincide (see
Figs,~4a,b). The asymptotic expressions in Eqs.~(\ref{eq46})
and (\ref{eq34}) with the coefficient (\ref{eq45}) are
applicable at $T<45\,$K and $T<35\,$K, respectively.
In Fig.~4a the asymptotic expression for the thermal
correction to the free energy achieves its maximal value at
about $T\approx 105\,$K. This is explained by the relatively large
static dielectric permittivity of Si.

Comparing all the above figures, one can conclude that the obtained
asymptotic expressions for the free energy (containing two
terms of order $\tau^3$ and $\tau^4$) have a wider
application range than the asymptotic expressions for the
pressure containing only one term of order $\tau^4$.
At the same time, by decreasing the separation distance one can
widen the range of temperatures where our asymptotic expressions
are applicable. The remarkable feature of Figs.~2--4 is
the monotonous increase of the magnitude of all thermal
corrections with the increase of temperature for real
dielectrics (see solid lines overlapping with our asymptotic
expressions in the applicability region of the latter).
This confirms the same conclusion made in
Refs.~\cite{R3,R4} based on qualitative thermodynamical
considerations.

\section{Role of the zero-frequency term in the Matsubara sum
for dielectrics}

According to Eq.~(\ref{eq1}), the free energy of the
van der Waals and Casimir interaction is represented by the
Matsubara sum from zero to infinity. The zero-frequency term
in Eq.~(\ref{eq1}),
\begin{equation}
{\cal F}_0(a,T)=\frac{k_BT}{16\pi a^2}
\int_{0}^{\infty}ydy\ln\left[1-
r_{\|}^{(1)}(0,y)r_{\|}^{(2)}(0,y)e^{-y}\right],
\label{eq53}
\end{equation}
\noindent
is of special interest.
For dielectrics with finite static dielectric
permittivities ${\ee}_0^{(k)}$ from Eq.~(\ref{eq2})
it follows:
\begin{equation}
r_{\|}^{(k)}(0,y)=\frac{{\ee}_0^{(k)}-1}{{\ee}_0^{(k)}+1},
\quad
r_{\|}^{(k)}(0,y)=0.
\label{eq54}
\end{equation}
\noindent
Eqs.~(\ref{eq53}) and (\ref{eq54}) were already used in
Eqs.~(\ref{eq17}) and (\ref{eq49}) to obtain the asymptotic
expressions for the free energy at high temperatures where the
contributions from all terms in the Matsubara sum with
$l\geq 1$ are exponentially small.

Note that the second equality in Eqs.~(\ref{eq54}) is somewhat
analogous with the same equality in the case of real metals
described by the Drude dielectric function \cite{B_S}.
In reality, however, for metals the second reflection
coefficient at zero frequency is $r_{\|}(0,y)=1$, i.e., equal
to its physical value. Together with $r_{\bot}(0,y)=0$,
this leads to the violation of the Nernst heat theorem for
perfect crystal lattices with no impurities \cite{R4} and to
contradictions with experiment \cite{Lam05,Lam_T,Decca1,Decca2,17}.
On the contrary, for dielectrics  $r_{\|}^{(k)}(0,y)$ in
Eq.~(\ref{eq54}) is larger than its physical value at normal
incidence [the latther is equal to
$(\sqrt{{\ee}_0^{(k)}}-1)/(\sqrt{{\ee}_0^{(k)}}+1)$] and,
as was demonstrated in Sec.~IV, the Lifshitz formula
incorporating Eqs.~(\ref{eq54}) is in perfect agreement with
the Nernst heat theorem.

We now turn to a problem of outstanding importance which
arises when one includes the dc conductivity of dielectric
materials into the model of their
dielectric response. What this means is that, instead of
dielectric permittivities ${\ee}_l^{(k)}={\ee}^{(k)}(i\xi_l)$,
one uses \cite{atFric,SurfSci}
\begin{equation}
\tilde{\ee}_l^{(k)}\equiv\tilde{\ee}^{(k)}(i\xi_l)
={\ee}^{(k)}(i\xi_l)+\frac{4\pi\sigma_0^{(k)}}{\xi_l}=
{\ee}_l^{(k)}+\frac{\beta^{(k)}(T)}{l}.
\label{eq55}
\end{equation}
\noindent
Here $\sigma_0^{(k)}$ is the dc conductivity of the plate
materials and $\beta^{(k)}=2\hbar\sigma_0^{(k)}/(k_BT)$.
It is common knowledge \cite{Slater} that the conductivity of
dielectrics depends on the temperature as
$\sigma_0^{(k)}\sim\exp(-b^{(k)}/T)$ where $b^{(k)}$ is
determined by the energy gap $\Delta\!^{(k)}$ which is
different for different materials.
It cannot be too highly stressed that for dielectrics the
conductivity at constant current is very low. To take
an example \cite{SiO2}, for SiO${}_2$ at $T=300\,$K
it holds $\beta^{(2)}\sim 10^{-12}$ and, thus, negligible
for all $l\geq 1$. From physical considerations the inclusion
of the term $\beta^{(k)}(T)/l$ into the model of the
dielectric response (\ref{eq55}) seems of dubious value
since $\beta^{(k)}(T)$ quickly decreases with decrease
of $T$ and, thus, remains negligible at any $T$.

In spite of this, the substitution of Eq.~(\ref{eq55}) into
the Lifshitz formula (\ref{eq1}) leads to Eq.~(\ref{eq53})
with different value of one of the
reflection coefficients at zero frequency than
in Eq.~(\ref{eq54}):
\begin{equation}
\tilde{r}_{\|}^{(k)}(0,y)=1,
\quad
\tilde{r}_{\|}^{(k)}(0,y)=0.
\label{eq56}
\end{equation}
\noindent
The change from Eq.~(\ref{eq54}) to Eq.~(\ref{eq56}) has
far-reaching consequences. To analyze them, we calculate the
free energy $\tilde{\cal F}(a,T)$ with the dielectric
permittivities $\tilde{\ee}_l^{(k)}$ identically
rearranging Eq.~(\ref{eq1}) to the form
\begin{eqnarray}
&&
\tilde{\cal F}(a,T)=\frac{k_BT}{16\pi a^2}
\int_{0}^{\infty}ydy\left\{
\vphantom{\ln\left[\frac{(\ezf -1)(\ezs -1)}{(\ezf +1)(\ezs +1)}
e^{-y}\right]}
\ln\left(1-e^{-y}\right)
\right.
\nonumber \\
&&\phantom{aa}
\left.
-\ln\left[1-\frac{(\ezf -1)(\ezs -1)}{(\ezf +1)(\ezs +1)}
e^{-y}\right]\right\}
\nonumber \\
&&\phantom{aa}
+\frac{k_BT}{16\pi a^2}
\int_{0}^{\infty}ydy
\ln\left[1-\frac{(\ezf -1)(\ezs -1)}{(\ezf +1)(\ezs +1)}
e^{-y}\right]
\label{eq57} \\
&&\phantom{aa}
+\frac{k_BT}{8\pi a^2}\sum\limits_{l=1}^{\infty}
\int_{\zeta_l}^{\infty}ydy\left\{
\ln\left[1-\tilde{r}_{\|}^{(1)}(\zeta_l,y)
\tilde{r}_{\|}^{(2)}(\zeta_l,y)e^{-y}\right]
\right.
\nonumber \\
&&\phantom{aa}
+\left.
\ln\left[1-\tilde{r}_{\bot}^{(1)}(\zeta_l,y)
\tilde{r}_{\bot}^{(2)}(\zeta_l,y)e^{-y}\right]
\right\},
\nonumber
\end{eqnarray}
\noindent
where the reflection coefficients $\tilde{r}_{\|,\bot}^{(k)}$
are obtained from Eq.~(\ref{eq2}) by replacing
${\ee}_l^{(k)}$ with $\tilde{\ee}_l^{(k)}$.

Now we expand the last, third term on the right-hand side
of Eq.~(\ref{eq57}) in powers of the small parameters
$\beta^{(k)}/l$. Combining the zero-order contribution in
this expansion with the second term on the right-hand side
of Eq.~(\ref{eq57}), one obtains the free energy
${\cal F}(a,T)$ calculated with the dielectric permittivities
${\ee}_l^{(k)}$. Calculating the first integral on the
right-hand side of Eq.~(\ref{eq57}), we arrive at
\begin{eqnarray}
&&
\tilde{\cal F}(a,T)= {\cal F}(a,T)
\label{eq58} \\
\phantom{aa}
&&
-\frac{k_BT}{16\pi a^2}\left\{\zeta(3)-
\mbox{Li}_3\left[\frac{(\ezf -1)(\ezs -1)}{(\ezf +1)(\ezs +1)}
e^{-y}\right]\right\}+R(a,T).
\nonumber
\end{eqnarray}
\noindent
In this formula $R(a,T)$ is of order $O(\beta^{(k)}/l)$
and it stands
for the first and higher-order contributions in the
expansion of the third term on the right-hand side
of Eq.~(\ref{eq57}) in powers of $\beta^{(k)}/l$.
The explicit expression for $R(a,T)$ is given in Eqs.~(\ref{c1})
and (\ref{c2}) of Appendix C. As is shown in Appendix C,
$R(a,T)$ exponentially goes to zero with the decrease of $T$.

Eq.~(\ref{eq58}) leads to the important conclusion about the
thermodynamic inconsistency of the Lifshitz formula for
dielectrics if one includes the dc conductivity
in the model of dielectric response. Substituting Eq.~(\ref{eq58})
into Eq.~(\ref{eq24}), we obtain the entropy for the plates with
the dielectric permittivities (\ref{eq55}) in the form
\begin{eqnarray}
&&
\tilde{S}(a,T)= {S}(a,T)
\label{eq59} \\
\phantom{aa}
&&
+\frac{k_B}{16\pi a^2}\left\{\zeta(3)-
\mbox{Li}_3\left[\frac{(\ezf -1)(\ezs -1)}{(\ezf +1)(\ezs +1)}
\right]\right\}-\frac{\partial R(a,T)}{\partial T},
\nonumber
\end{eqnarray}
\noindent
where $S(a,T)$ is the entropy for the plates with the dielectric
permittivities ${\ee}^{(k)}$ given by Eq.~(\ref{eq48}).

In the limit $T\to 0$ it follows:
\begin{equation}
\tilde{S}(a,0)=\frac{k_B}{16\pi a^2}\left\{\zeta(3)-
\mbox{Li}_3\left[\frac{(\ezf -1)(\ezs -1)}{(\ezf +1)(\ezs +1)}
\right]\right\}>0.
\label{eq60}
\end{equation}
\noindent
Eq.~(\ref{eq60}) depends on the parameter of the system under
consideration (the separation distance $a$) and implies
the violation of the Nernst heat theorem. Thus, the dc conductivity
of a dielectric is irrelevant to the
origin of the van der Waals and Casimir forces and must not
be included in the models of the dielectric response. The neglect
of this rule results in the violation of thermodynamics.
Physically it is amply clear that for high-frequency phenomena
like the van der Waals and Casimir forces the behavior of
dielectric materials at low frequencies is described by the
static dielectric permittivities (the approach used by
E.~M.~Lifshitz and his collaborators \cite{D_L_P}).
The results of this section provide the necessary
theoretical background for this conclusion.

\section{Conclusions and discussion}

In the foregoing, we have reconsidered the Lifshitz theory of the
Casimir and van der Waals interaction between dielectrics on the
basis of thermal quantum field theory in Matsubara formulation.
As was shown above, there are field theoretical  derivations of
the Lifshitz formula  without use of the fluctuation-dissipation
theorem which permit generalization for the presence
of dissipation. The special analysis demonstrates, however, that
only a particular
dissipation is compatible with the Lifshitz formula, those
without net heat losses and with balanced processes of
absorption and emission.

In this paper, we have analytically solved the long-standing
problem of the low-temperature (short separation) behavior of
the thermal corrections to the Casimir energy and pressure
between dissimilar dielectric plates and demonstrated that the
Casimir (van der Waals) entropy vanishes when the temperature
goes to zero. This was done, first, using the idealized model
of dilute dielectrics, and, then, for real dielectrics with
finite static dielectric permittivities. The free energy,
pressure and entropy of both the van der Waals and Casimir
interactions between dielectrics demonstrate at low
temperatures the same universal temperature dependence which
was previously discovered for ideal metals. This proves the
thermodynamic consistency of the Lifshitz theory in application
to dielectric plates. The obtained asymptotic expressions were
compared with the results of numerical computations for both
similar and dissimilar real dielectrics (silicon and vitreous
silica), and were found to be in excellent agreement.

Special attention was paid to the role of the zero-frequency
term in Matsubara sum in the case of dielectric plates and to
the extrapolation of the dielectric permittivity along the
imaginary frequency axis to zero frequency. As was proved above,
the inclusion of the conductivity at constant current in the model
of dielectric response leads to a modification of the
zero-frequency   term of the Lifshitz formula and to the violation
of the Nernst heat theorem. The conclusion was made that the
conductivity of dielectrics at constant current (low but
formally nonzero at nonzero temperature) is irrelevant to
physical phenomena described by the Lifshitz theory. This
conclusion leads to far-reaching consequences for both the
problem of noncontact atomic friction and for Casimir
interaction between real metals at nonzero temperature.

As to the problem of atomic friction, the observed effect is
many orders of magnitude larger \cite{30} than the van der Waals
friction between metals caused by the vacuum fluctuations and
thermal photons (in the experiment of Ref.~\cite{30} the
gold-coated tip and substrate were used). In
Refs.~\cite{atFric,SurfSci}
it has been proposed that measurements performed on metal films
are strongly affected by the underlying dielectric substrate.
To calculate the friction due to the substrate,
Refs.~\cite{atFric,SurfSci} use the Lifshitz-type formula including
the low dc conductivity in the model of dielectric
response. Then, the large value of friction coefficient is obtained
in rough agreement with the experimental data of Ref.~\cite{30}.
According to our results, the proposition of
Refs.~\cite{atFric,SurfSci} would not be in agreement with the
Nernst heat theorem. It is also qualitatively clear that the
low-frequency behavior of the dielectric permittivity in the region
below $\sim 600\,$rad/s cannot cause the large friction effect at
characteristic frequencies  from $3.75\times 10^{14}\,$rad/s to
$1.5\times 10^{17}\,$rad/s (in the experiment of Ref.\cite{30}
separations vary  from 1 to 400 nm).

Regarding  the application of the obtained results to real metals,
the case of two dielectric plates with including the dc conductivity
turns out to be analogous to two metal plates
described by the Drude dielectric function. In both cases the
region of low frequencies contributes significantly to the
description of high-frequency phenomena resulting in contradictions
with thermodynamics (see Introduction). It is well known that in
the region of the normal skin effect, described by the Drude
dielectric function, dissipation leads to the heating of a metal
(see Ref.~\cite{GKM_03} for details). For dissipation of this
kind, as discussed above, the field theoretical derivation of
the Lifshitz formula is inapplicable. In this connection it is not
surprising  also that the results obtained from the Lifshitz formula
combined with the Drude model are in conflict with the conclusions
obtained for ideal metals using the thermal quantum field theory
approach
with the Dirichlet boundary conditions. The proposed measurement of
the Casimir force between both metals and semiconductors at large
separations \cite{Lam_B} is aimed to resolve this contradiction
experimentally. At large separations (high temperatures) the
classical limit of the Casimir effect is achieved where  forces
acting between real and ideal metals practically coincide
\cite{F_M_R,Jaffe}.
In our opinion, this important experiment could
bring the final verification of the results obtained from the thermal
quantum field theory approach (see also Refs.~\cite{lowT,Napoli}
where two short-separation experiments are proposed with the same
aim).

It is worthwhile to note that in real experiments the most frequent
configuration is not the two parallel plates but a large sphere
above a plate. There is a prediction in recent literature
\cite{Jaffe} that for ideal metals in the case of open geometries
the thermal correction to the Casimir force is not as suppressed
as in the parallel plate geometry in the limit of zero
temperature. By this reason it is of much interest to derive
on fundamental grounds the low-temperature behavior of the
Casimir free energy between real bodies (metallic or dielectric)
for any open geometry.

To conclude, the above reconsideration of the Casimir and van der
Waals interactions between dielectric semispaces not only confirms the
mutual agreement  between the thermal quantum field theory approach,
Lifshitz formula and thermodynamics, but also suggests ways
for the resolution of other complicated problems in modern
applications of quantum electrodynamics.

\section*{ACKNOWLEDGMENTS}

G.L.K. and V.M.M. are grateful to the Center of Theoretical Studies and
the Institute for Theoretical
Physics, Leipzig University for their kind hospitality.
This work was supported by Deutsche Forschungsgemeinschaft grant
436\,RUS\,113/789/0-1. G.L.K. and V.M.M. were also partially
supported by Finep (Brazil).

\section*{APPENDIX A}
\setcounter{equation}{0}
\renewcommand{\theequation}{A\arabic{equation}}

In this Appendix we prove Eq.~(\ref{eq32}) in Sec.~IV. Let us present
Eq.~(\ref{eq28}) in the form
\begin{equation}
f(\zeta,y)=f_{\|}(\zeta,y)+f_{\bot}(\zeta,y),
\label{a1}
\end{equation}
\noindent
where
\begin{equation}
f_{\|,\bot}(\zeta,y)=y\ln\left[1-r_{\|,\bot}^{(1)}(\zeta,y)
r_{\|,\bot}^{(2)}(\zeta,y)e^{-y}\right].
\label{a2}
\end{equation}
\noindent
It is easy to check that $f_{\bot}(\zeta,y)$ does not contribute to the
first term on the right-hand side of Eq.~(\ref{eq32}).
Substituting Eq.~(\ref{eq31}) in $f_{\|}(\zeta,y)$ and expanding in
powers of $x=\tau t$ one obtains
\begin{eqnarray}
&&
f_{\|,\bot}(x,y)=y\ln\left[1-r_{0}^{(1)}
r_{0}^{(2)}e^{-y}\right]
\nonumber \\
&&
\phantom{a}
+\frac{\ezf+\ezs+2\ezf \ezs}{\left(\ezf +1\right)
\left(\ezs +1\right)}\,
\frac{\rf \rs e^{-y}}{y\left(1-\rf \rs e^{-y}\right)}\,x^2
\label{a3} \\
&&
\phantom{a}
+2\left[\frac{1}{\left(\ezf+1\right)}\,\frac{\xi_c^2}{{\omega^{(1)}}^2}+
\frac{1}{\left(\ezs +1\right)}\,\frac{\xi_c^2}{{\omega^{(2)}}^2}
\right]\,\frac{\rf \rs ye^{-y}}{1-\rf \rs e^{-y}}\,x^2
+O(x^3),
\nonumber
\end{eqnarray}
\noindent
where the static values of the reflection coefficients are found
from Eq.~(\ref{eq2})
\begin{equation}
r_0^{(k)}\equiv r_{\|}^{(k)}(0,y)=
\frac{\ee_0^{(k)}-1}{\ee_0^{(k)}+1}< 1.
\label{a4}
\end{equation}
\noindent
Note that for simplicity we consider only one oscillator in
Eq.~(\ref{eq31}) and omit index $j$. The obtained results,
however, are valid for any number of oscillators.

As a next step, we should integrate Eq.~(\ref{a3}) according to
Eq.~(\ref{eq30}). Integration of the first term on the right-hand
side of Eq.~(\ref{a3}) results in
\begin{eqnarray}
&&
Z_1(x)\equiv\int_x^{\infty}ydy
\ln\left(1-\rf \rs e^{-y}\right)
\nonumber \\
&&\phantom{aaa}
=-\sum\limits_{n=1}^{\infty}
\frac{(1+nx)e^{-nx}}{n^3}\left(\rf \rs\right)^n.
\label{a5}
\end{eqnarray}
\noindent
Expanding (\ref{a5}) in powers of $x$ and summing the obtained
series, we arrive at
\begin{eqnarray}
&&
Z_1(x)=-\mbox{Li}_3\left(\rf \rs\right)
\nonumber \\
&&\phantom{aaa}
-\frac{x^2}{2}\ln\left(1-\rf \rs\right)+O(x^3).
\label{a6}
\end{eqnarray}

Integration of the second term on the right-hand
side of Eq.~(\ref{a3}) contains the integral
\begin{eqnarray}
&&
Z_2(x)=x^2\int_x^{\infty}dy
\frac{\rf \rs e^{-y}}{y\left(1-\rf \rs e^{-y}\right)}
\nonumber \\
&&\phantom{aaa}
=-x^2\sum\limits_{n=1}^{\infty}
\left(\rf \rs\right)^n\mbox{Ei}(-nx).
\label{a7}
\end{eqnarray}

Integration of the third term on the right-hand
side of Eq.~(\ref{a3}) leads to the integral
\begin{eqnarray}
&&
Z_3(x)=x^2\int_x^{\infty}dy\,y
\frac{\rf \rs e^{-y}}{1-\rf \rs e^{-y}}
\nonumber \\
&&\phantom{aaa}
=x^2\sum\limits_{n=1}^{\infty}
\left(\rf \rs\right)^n\frac{(1+nx)e^{-nx}}{n^2}.
\label{a8}
\end{eqnarray}
\noindent
Expanding in powers of $x$ in Eq.~(\ref{a8}) and performing
the summation, we obtain
\begin{equation}
Z_3(x)=x^2\mbox{Li}_2\left(\rf \rs\right)
-\frac{x^4}{2}\frac{\rf \rs}{1-\rf \rs}+O(x^5).
\label{a9}
\end{equation}

Let us now calculate the different contributions to
the quantity (\ref{eq32}). According to Eq.~(\ref{a6})
\begin{equation}
Z_1(ix)-Z_1(-ix)=O(x^3).
\label{a10}
\end{equation}
\noindent
From Eq.~(\ref{a7}) it follows:
\begin{equation}
Z_2(ix)-Z_2(-ix)=i\pi x^2\frac{\rf \rs}{1-\rf \rs}+O(x^3).
\label{a11}
\end{equation}
\noindent
Finally, Eq.~(\ref{a9}) leads to
\begin{equation}
Z_3(ix)-Z_3(-ix)=O(x^5).
\label{a12}
\end{equation}
\noindent
As is seen from Eqs.~(\ref{a10})--(\ref{a12}), only
Eq.~(\ref{a11}) contributes to the leading term on the
right-hand side of Eq.~(\ref{eq32}). As to the frequency
dependence of the dielectric permittivity in accordance
to Eq.~(\ref{eq31}), it contributes only to the fifth order
term [see the last term on the right-hand side of
Eq.~(\ref{a3}) and Eq.~(\ref{a12})].

Assigning the numerical coefficient to $Z_2$ as in the
second term on the right-hand side of Eq.~(\ref{a3}) and
using Eq.~(\ref{a11}), we arrive at the result
\begin{equation}
F(ix)-F(-ix)=\frac{\ezf +\ezs +2\ezf \ezs }{(\ezf +1)(\ezs +1)}
\,\frac{i\pi \rf \rs}{1-\rf \rs}x^2+O(x^3).
\label{a13}
\end{equation}
\noindent
In terms of a notation (\ref{a4}), this coincides with
Eq.~(\ref{eq32}).

\section*{APPENDIX B}
\setcounter{equation}{0}
\renewcommand{\theequation}{B\arabic{equation}}

This Appendix is devoted to the derivation of Eq.~(\ref{eq43})
in Sec.~IV where the function $\Phi_{\|}(ix)$ was used in the
calculation of pressure.
As was shown in Appendix A, the dependence of the dielectric
permittivity on frequency contributes to the expansion of $F(ix)$
in powers of $x$, used in the calculation of the free energy, starting
from only the 5th power [i.e., to $\Phi_{\|}(ix)$ starting from
the 4th power in $x$]. Here we are looking for the lowest (third)
order expansion term of $\Phi_{\|}(ix)$. Because of this, it is
possible to disregard the frequency dependence of $\ee $ and
describe the plate materials by their static dielectric
permittivities.

Let us identically rearrange Eq.~(\ref{eq37}) by subtracting and
adding the two first expansion terms of the function under the
integral in powers of $x$
\begin{eqnarray}
&&
\Phi_{\|}(x)=\int_x^{\infty}dy
\left[y^2\frac{r_{\|}^{(1)}(x,y)r_{\|}^{(2)}(x,y)e^{-y}}{1-
r_{\|}^{(1)}(x,y)r_{\|}^{(2)}(x,y)e^{-y}}-
y^2\frac{\rf \rs e^{-y}}{1-\rf \rs e^{-y}}
\right.
\nonumber \\
&&\phantom{aa}
\left.
+x^2\frac{\ezf +\ezs +2\ezf \ezs }{(\ezf +1)(\ezs +1)}\,
\frac{\rf \rs e^{-y}}{\left(1-\rf \rs e^{-y}\right)^2}\right]
\label{b1} \\
&&
\phantom{aa}
+\int_x^{\infty}y^2dy\frac{\rf \rs e^{-y}}{1-\rf \rs e^{-y}}
-x^2\frac{\ezf +\ezs +2\ezf \ezs }{(\ezf +1)(\ezs +1)}\,
\int_x^{\infty}dy
\frac{\rf \rs e^{-y}}{\left(1-\rf \rs e^{-y}\right)^2}.
\nonumber
\end{eqnarray}

We consider the first integral on the right-hand side of
Eq.~(\ref{b1}) written in terms of a new variable $v=y/x$
\begin{eqnarray}
&&
Q_1(x)=x^3\int_1^{\infty}dv
\left[v^2\frac{r_{\|}^{(1)}(0,v)r_{\|}^{(2)}(0,v)e^{-vx}}{1-
r_{\|}^{(1)}(0,v)r_{\|}^{(2)}(0,v)e^{-vx}}-
v^2\frac{\rf \rs e^{-vx}}{1-\rf \rs e^{-vx}}
\right.
\nonumber \\
&&\phantom{aa}
\left.
+\frac{\ezf +\ezs +2\ezf \ezs }{(\ezf +1)(\ezs +1)}\,
\frac{\rf \rs e^{-vx}}{\left(1-\rf \rs e^{-vx}\right)^2}\right],
\label{b2}
\end{eqnarray}
\noindent
where, in accordance with Eq.~(\ref{eq2}),
\begin{equation}
r_{\|}^{(k)}(0,v)=\frac{{\ee}_0^{(k)}v-
\sqrt{v^2+{\ee}_0^{(k)}-1}}{{\ee}_0^{(k)}v+
\sqrt{v^2+{\ee}_0^{(k)}-1}}.
\label{b3}
\end{equation}

The leading expansion order of $Q_1(x)$ from  Eq.~(\ref{b2})
in powers of $x$ is
\begin{eqnarray}
&&
x^3\int_1^{\infty}dv
\left[v^2\frac{r_{\|}^{(1)}(0,v)r_{\|}^{(2)}(0,v)}{1-
r_{\|}^{(1)}(0,v)r_{\|}^{(2)}(0,v)}-
v^2\frac{\rf \rs }{1-\rf \rs }
\right.
\nonumber \\
&&\phantom{aa}
\left.
+\frac{\ezf +\ezs +2\ezf \ezs }{(\ezf +1)(\ezs +1)}\,
\frac{\rf \rs }{\left(1-\rf \rs \right)^2}\right].
\label{b4}
\end{eqnarray}
\noindent
The explicit calculation of the integral in Eq.~(\ref{b4})
leads to
\begin{eqnarray}
&&
Q_1(x)=\left[1-\frac{\ezf +\ezs}{\sqrt{\ezf}+\sqrt{\ezs}}-
\frac{3 \ezf \ezs \sqrt{\ezf \ezs}}{\left(\sqrt{\ezf}+
\sqrt{\ezs}\right)\left(\ezf +\ezs \right)}\right.
\label{b5} \\
&&\phantom{a}
+\frac{{\ezf}^2 {\ezs}^2\left(3\sqrt{\ezf \ezs}+
\ezf +\ezs \right)}{\left(\sqrt{\ezf}+
\sqrt{\ezs}\right)\left(\ezf +\ezs \right)^2}-
\frac{{\ezf} {\ezs}\left(\sqrt{\ezf \ezs}-
\ezf -\ezs \right)}{\left(\sqrt{\ezf}+
\sqrt{\ezs}\right)\left(\ezf +\ezs \right)^2}
\nonumber \\
&&\phantom{a}
-\frac{\sqrt{\ezf \ezs}\left(
\ezf -\ezs \right)^2}{\left(\sqrt{\ezf}+
\sqrt{\ezs}\right)\left(\ezf +\ezs \right)^2}
\nonumber \\
&&\phantom{a}\left.
-
\frac{3{\ezf}^2 {\ezs}^2\left(\ezf -1\right)\left(\ezs -
1\right)}{\left(\ezf -\ezs\right)\left(\ezf +\ezs\right)^{5/2}}
\mbox{Artanh}\frac{\sqrt{\ezf +\ezs}\left(\sqrt{\ezf} -
\sqrt{\ezs}\right)}{\sqrt{\ezf \ezs}-\ezf -\ezs}\right]\,
\frac{x^3}{6}
\nonumber \\
&&\phantom{a}
+\frac{x^3}{3}\,\frac{\rf \rs }{1-\rf \rs }-
x^3\frac{\ezf +\ezs +2\ezf \ezs }{(\ezf +1)(\ezs +1)}\,
\frac{\rf \rs }{\left(1-\rf \rs \right)^2}+O(x^4).
\nonumber
\end{eqnarray}

Now we calculate the second integral on the right-hand
side of Eq.~(\ref{b1}),
\begin{eqnarray}
&&
Q_2(x)=\int_x^{\infty}y^2dy\frac{\rf \rs e^{-y}}{1-\rf \rs e^{-y}}
\label{b6} \\
&&\phantom{a}
=\sum\limits_{n=1}^{\infty}\left(\rf \rs\right)^n
\frac{(2+2nx+n^2x^2)e^{-nx}}{n^3}.
\nonumber
\end{eqnarray}
\noindent
Expanding Eq.~(\ref{b6}) in powers of $x$, one obtains
\begin{equation}
Q_2(x)=2\mbox{Li}_3\left(\rf \rs\right)-
\frac{x^3}{3}\,\frac{\rf \rs }{1-\rf \rs }+O(x^4).
\label{b7}
\end{equation}

The integral contained in the third term  on the right-hand
side of Eq.~(\ref{b1}) is simply calculated,
\begin{eqnarray}
&&
Q_3(x)=x^2
\int_x^{\infty}dy
\frac{\rf \rs e^{-y}}{\left(1-\rf \rs e^{-y}\right)^2}
\label{b8} \\
&&\phantom{a}
=x^2\frac{\rf \rs e^{-x}}{1-\rf \rs e^{-x}}.
\nonumber
\end{eqnarray}
It can be expanded in powers of $x$ as follows:
\begin{equation}
Q_3(x)=x^2\frac{\rf \rs }{1-\rf \rs }-
x^3\frac{\rf \rs }{\left(1-\rf \rs \right)^2}+O(x^4).
\label{b9}
\end{equation}

Finally, according to Eq.~(\ref{b1}), the function under
consideration is given by
\begin{equation}
\Phi_{\|}(x)=Q_1(x)+Q_2(x)-
\frac{\ezf +\ezs +2\ezf \ezs }{(\ezf +1)(\ezs +1)}\,Q_3(x),
\label{b10}
\end{equation}
\noindent
where $Q_1(x)$, $Q_2(x)$, and $Q_3(x)$ are found in
Eqs.~(\ref{b5}), (\ref{b7}), and (\ref{b9}), respectively.
It is notable that the contributions of the third power
from $Q_2(x)$ and $Q_3(x)$ in Eq.~(\ref{b10}) cancel
the second and third contributions from $Q_1(x)$.

Using Eq.~(\ref{b10}), one arrives at Eq.~(\ref{eq43})
after some identical rearrangements.

\section*{APPENDIX C}
\setcounter{equation}{0}
\renewcommand{\theequation}{C\arabic{equation}}

The quantity $R(a,T)$ was introduced in Eq.~(\ref{eq58})
of Sec.~VI and has the
following explicit form:
\begin{equation}
R(a,T)=R^{(1)}(a,T)+R^{(2)}(a,T)+\mbox{\large$O$}\left[\left(
{\beta^{(k)}}/{l}\right)^2\right],
\label{c1}
\end{equation}
\noindent
where
\begin{eqnarray}
&&
R^{(1)}(a,T)=\frac{k_BT}{8\pi a^2}
\sum\limits_{l=1}^{\infty}\left\{
\frac{\beta^{(1)}}{l}\int_{\zeta_l}^{\infty}
\frac{dy\,y^2e^{-y}}{\sqrt{y^2+\zeta_l^2({\ee}_l^{(1)}-1)}}
\right.
\nonumber \\
&&\phantom{aa}
\times\left[\frac{(2-{\ee}_l^{(1)})\zeta_l^2-
2y^2}{({\ee}_l^{(1)}y+\sqrt{y^2+\zeta_l^2({\ee}_l^{(1)}-1)})^2}\,\,
\frac{r_{\|}^{(2)}(\zeta_l,y)}{1-
r_{\|}^{(1)}(\zeta_l,y)r_{\|}^{(2)}(\zeta_l,y)e^{-y}}
\right.
\label{c2} \\
&&\phantom{aa}\left.\left.
-\frac{\zeta_l^2}{\sqrt{y^2+\zeta_l^2({\ee}_l^{(1)}-1)}+y}\,\,
\frac{r_{\bot}^{(2)}(\zeta_l,y)}{1-
r_{\bot}^{(1)}(\zeta_l,y)r_{\bot}^{(2)}(\zeta_l,y)e^{-y}}
\right]\right\}
\nonumber
\end{eqnarray}
\noindent
and $R^{(2)}(a,T)$ is obtained from $R^{(1)}(a,T)$ by
interchanging of the upper indices (1) and (2).
In this Appendix we demonstrate that $R(a,T)$ vanishes
exponentially when $T\to 0$.

Let us consider the integral with respect to $y$ from
Eq.~(\ref{c2}), expand the integrated function in powers of
$\tau$ (we recall that $\zeta_l=\tau l$) and restrict
ourselves by the main contribution to the result at
$\tau =0$:
\begin{equation}
-2\int_{\zeta_l}^{\infty}dy\frac{ye^{-y}}{(\ezf +1)^2}\,
\frac{\rs}{1-\rf \rs e^{-y}},
\label{c3}
\end{equation}
\noindent
where $r_0^{(k)}$ was defined in Eq.~(\ref{a4}).
With account of Eq.~(\ref{a4}), Eq.~(\ref{c3}) can be
rearranged to the form
\begin{eqnarray}
&&
-\frac{2}{({\ezf}^2-1)}
\int_{\zeta_l}^{\infty}dy\,y
\frac{\rf \rs e^{-y}}{1-\rf \rs e^{-y}}
\label{c5} \\
&&\phantom{aa}=
-\frac{2}{({\ezf}^2-1)}\sum\limits_{n=1}^{\infty}
\left(\rf \rs \right)^n
\int_{\zeta_l}^{\infty}dy\,ye^{-ny}
\nonumber \\
&&\phantom{aa}=
-\frac{2}{({\ezf}^2-1)}\sum\limits_{n=1}^{\infty}
\left(\rf \rs \right)^n
\frac{1+n\zeta_l}{n^2}e^{-n\zeta_l}.
\nonumber
\end{eqnarray}
\noindent
Substituting Eq.~(\ref{c5}) in Eq.~(\ref{c2}), we find
\begin{eqnarray}
&&
R^{(1)}(a,T)=
-\frac{k_BT\beta^{(1)}}{4\pi a^2\left({\ezf}^2-1\right)}
\sum\limits_{n=1}^{\infty}
\frac{\left(\rf \rs \right)^n}{n^2}
\nonumber \\
&&\phantom{aa}
\times\left[\sum\limits_{l=1}^{\infty}\frac{e^{-n\tau l}}{l}
+n\tau\sum\limits_{l=1}^{\infty}e^{-n\tau l}\right]
\nonumber \\
&&\phantom{aa}
=-\frac{k_BT\beta^{(1)}}{4\pi a^2\left({\ezf}^2-1\right)}
\sum\limits_{n=1}^{\infty}
\frac{\left(\rf \rs \right)^n}{n^2}
\label{c6} \\
&&\phantom{aa}
\times\left[-\ln\left(1-e^{-n\tau}\right)+
\frac{n\tau}{e^{n\tau}-1}\right].
\nonumber
\end{eqnarray}
\noindent
It is readily seen that
\begin{equation}
-\ln\left(1-e^{-n\tau}\right)+
\frac{n\tau}{e^{n\tau}-1}=-\ln\tau+1-\ln n+O(\tau^2).
\label{c7}
\end{equation}

The substitution of the leading term on the right-hand side
of Eq.~(\ref{c7}) in Eq.~(\ref{c6}) results in
\begin{eqnarray}
&&
R^{(1)}(a,T)=
\frac{k_BT\beta^{(1)}\ln\tau}{4\pi a^2\left({\ezf}^2-1\right)}
\sum\limits_{n=1}^{\infty}
\frac{\left(\rf \rs \right)^n}{n^2}+T\beta^{(1)}O(\tau^0)
\nonumber \\
&&\phantom{aa}
=\frac{k_B\mbox{Li}_2\left(\rf \rs \right)}{4\pi a^2\left({\ezf}^2
-1\right)}
T\beta^{(1)}\ln\tau+T\beta^{(1)}O(\tau^0).
\label{c8}
\end{eqnarray}
\noindent
Taking into account that
$\beta^{(1)}\sim(1/T)\exp(-b^{(1)}/T)$ (see Sec.~VI), we get the
conclusion that the temperature dependence of $R^{(1)}(a,T)$ is
determined by the term
\begin{equation}
R^{(1)}(a,T)\sim e^{-b^{(1)}/T}\ln T,
\label{c9}
\end{equation}
\noindent
i.e., both $R^{(1)}(a,T)$ and its derivative go to zero as
$\exp(-b^{(1)}/T)$ when the temperature goes to zero.

In perfect analogy to $R^{(1)}(a,T)$, the same conclusion is
obtained for $R^{(2)}(a,T)$. The terms of the second and
higher powers in $\beta^{(k)}$ [see Eq.~(\ref{c1})] go to
zero even faster than $R^{(k)}(a,T)$ when $T\to 0$.
In this way, with account of Eq.~(\ref{c1}), we have proven
the exponentially fast vanishing of $R(a,T)$ with
decrease of the temperature.


\begin{figure*}
\vspace*{-8cm}
\includegraphics{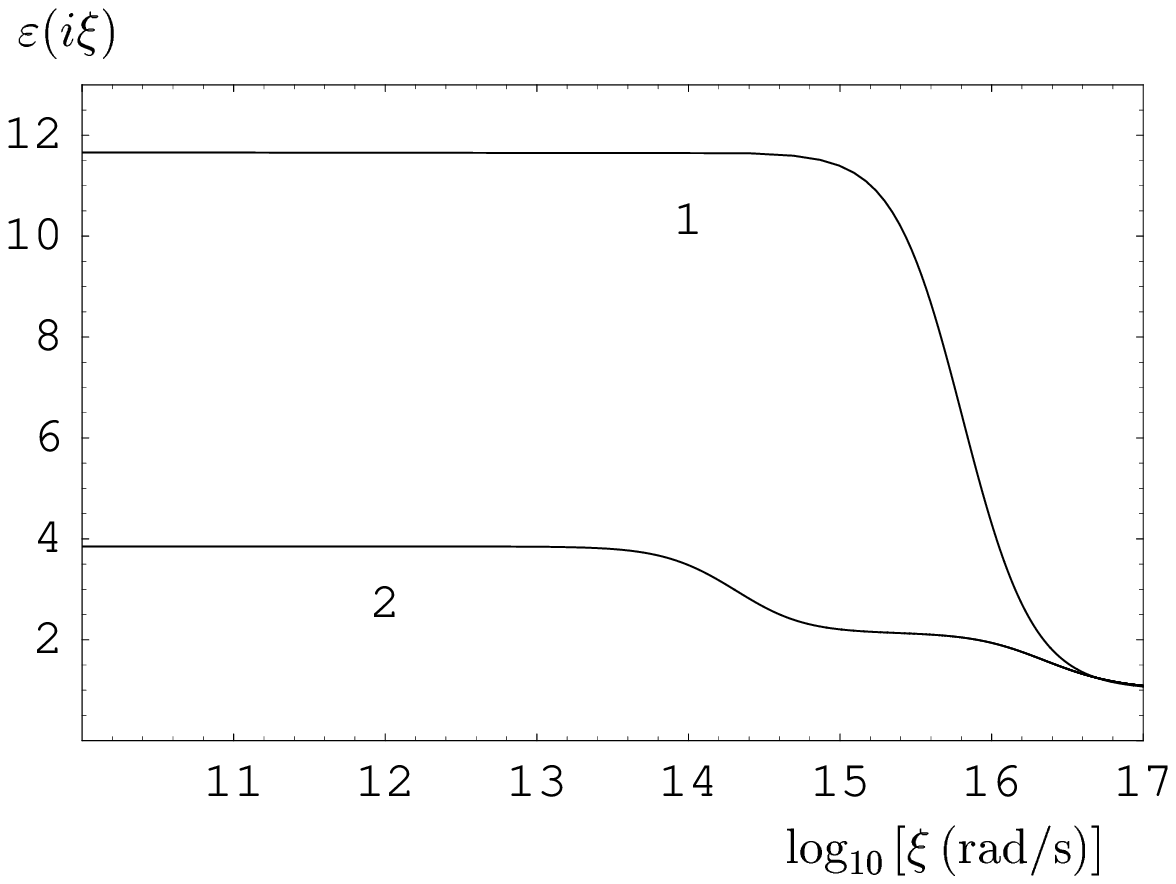}
\vspace*{-9cm}
\caption{
Dielectric permittivity of Si (line 1) and vitreous
$\mbox{SiO}_2$ (line 2) along
the imaginary frequency axis as a function of the logarithm
of frequency.
}
\end{figure*}
\begin{figure*}
\vspace*{-2cm}
\includegraphics{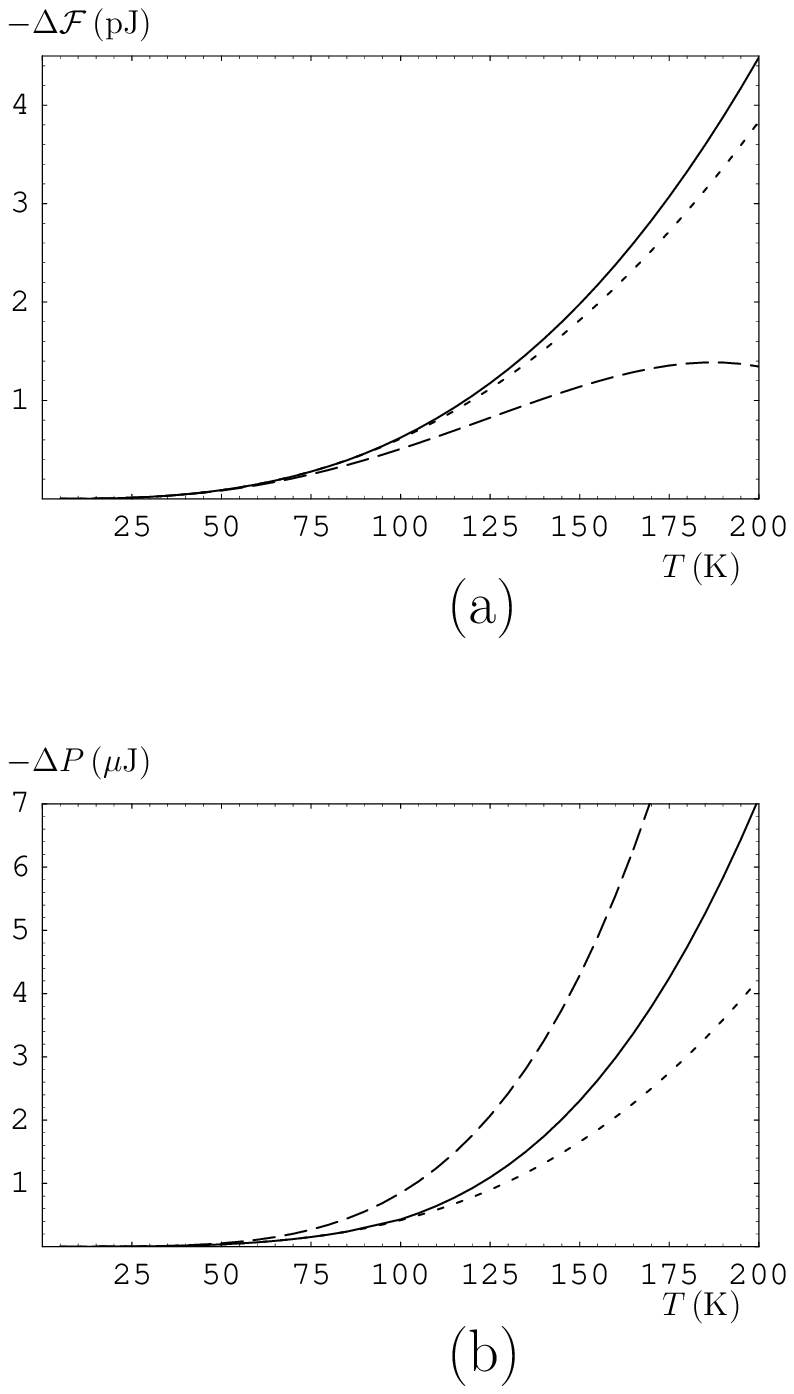}
\vspace*{-7cm}
\caption{Magnitudes of the thermal corrections to the energy (a)
and pressure (b) in configuration of two plates one made of
Si and another one of $\mbox{SiO}_2$  at a separation
$a=400\,$nm as a function of temperature calculated by the
use of different approaches: by the Lifshitz formula and
tabulated optical data (solid lines), by the Lifshitz formula
and static dielectric permittivities (short-dashed lines),
by the asymptotic expressions in Eqs.~(\ref{eq33}),
(\ref{eq34}) and (\ref{eq44}) (long-dashed lines).
}
\end{figure*}
\begin{figure*}
\vspace*{-2cm}
\includegraphics{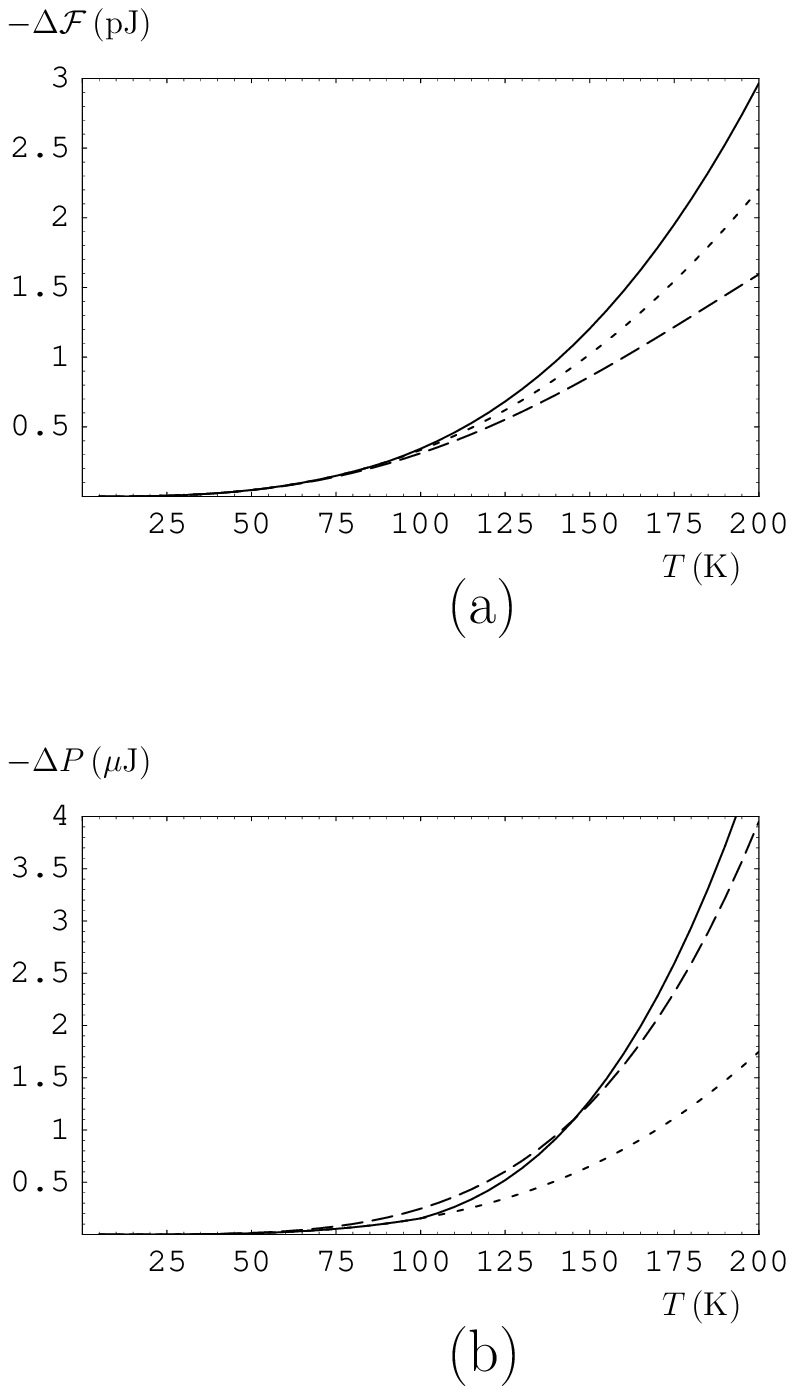}
\vspace*{-7cm}
\caption{Magnitudes of the thermal corrections to the energy (a)
and pressure (b) in configuration of two plates made of
vitreous $\mbox{SiO}_2$  at a separation
$a=450\,$nm as a function of temperature calculated by the
use of different approaches: by the Lifshitz formula and
tabulated optical data (solid lines), by the Lifshitz formula
and static dielectric permittivities (short-dashed lines),
by the asymptotic expressions in Eqs.~(\ref{eq34}),
(\ref{eq45}) and (\ref{eq46}) (long-dashed lines).
}
\end{figure*}
\begin{figure*}
\vspace*{-2cm}
\includegraphics{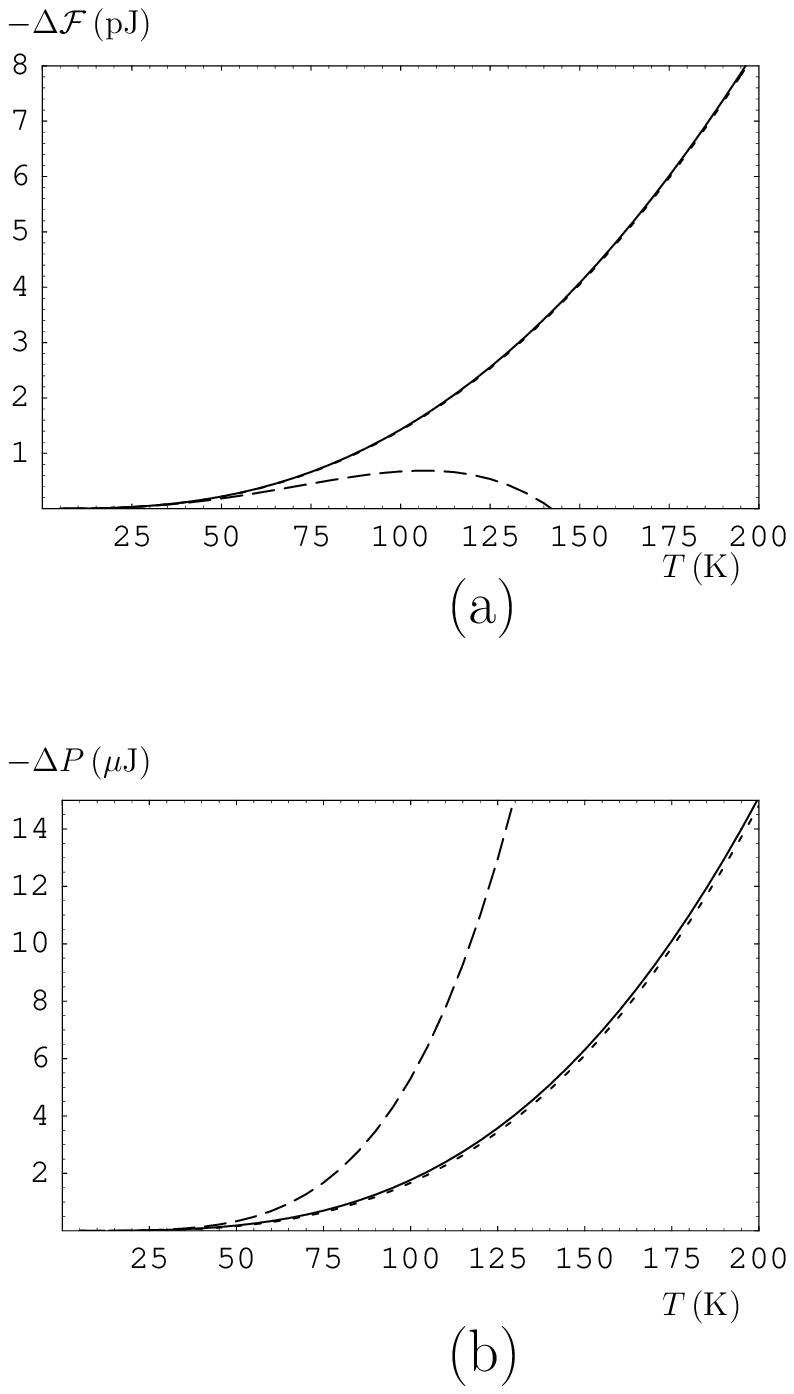}
\vspace*{-7cm}
\caption{Magnitudes of the thermal corrections to the energy (a)
and pressure (b) in configuration of two plates  made of
Si at a separation
$a=300\,$nm as a function of temperature calculated by the
use of different approaches: by the Lifshitz formula and
tabulated optical data (solid lines), by the Lifshitz formula
and static dielectric permittivities (short-dashed lines),
by the asymptotic expressions in Eqs.~(\ref{eq34}),
(\ref{eq45}) and (\ref{eq46}) (long-dashed lines).
}
\end{figure*}
\end{document}